\documentclass[twocolumn, twocolappendix]{aastex631}

\usepackage{amsmath}
\usepackage{graphicx}
\usepackage{url}

\newcommand{\NUPA}{\affiliation{ Department of Physics \& Astronomy, Northwestern University, Evanston, IL 60208, USA}}
\newcommand{\NUCIERA}{\affiliation{Center for Interdisciplinary Exploration \& Research in Astrophysics (CIERA), Northwestern University, Evanston, IL 60208, USA}}

\newcommand{\msun}{\mathrm{M}_\odot}

\newcommand{\vinf}{v_\mathrm{inf}}
\newcommand{\vorb}{v_\mathrm{orb}}
\newcommand{\vcrit}{v_\mathrm{crit}}
\newcommand{\vkick}{v_\mathrm{kick}}
\newcommand{\vesc}{v_\mathrm{esc}}

\newcommand{\vgw}{v_\mathrm{gw}}
\newcommand{\sigdisp}{\sigma_\mathrm{disp}}

\newcommand{\bmax}{b_\mathrm{max}}
\newcommand{\rpmax}{r_\mathrm{p,max}}

\newcommand{\tenc}{T_\mathrm{enc}}

\newcommand{\tgw}{T_\mathrm{GW}}

\newcommand{\tfinal}{T_\mathrm{final}}
\newcommand{\tint}{T_\mathrm{int}}
\newcommand{\tmin}{T_\mathrm{min}}

\newcommand{\crosssec}[1]{\sigma_\mathrm{#1}}
\newcommand{\rate}[1]{\Sigma_\mathrm{#1}}
\newcommand{\prob}[1]{P_\mathrm{#1}}

\newcommand{\Deltaave}{\langle\Delta\rangle}
\newcommand{\dd}{\mathrm d}
\newcommand{\nfid}{n=10^6\,\mathrm{pc}^{-3}}

\begin{document}

\title{Survival Analysis of Intermediate-Mass Black Holes in Dense Star Clusters}

\correspondingauthor{Miguel A. S. Martinez}
\email{miguelmartinez2025@u.northwestern.edu}

\author[0000-0001-5285-4735]{Miguel A. S. Martinez}
\NUPA
\NUCIERA

\author[0000-0002-0933-6438]{Elena Gonz\'{a}lez Prieto}
\NUPA
\NUCIERA

\author[0000-0002-7132-418X]{Frederic A. Rasio}
\NUPA
\NUCIERA

\begin{abstract}

Recently, an intermediate-mass black hole (IMBH) candidate was announced in the Galactic globular cluster Omega Centauri. 
IMBHs at the lower end of the traditional mass range have also been detected through gravitational-wave transients, though their formation and subsequent growth linking the two mass scales remains a mystery. 
One way IMBHs may be produced is through the collapse of very massive stars produced by stellar collisions in dense stellar environments. 
However, IMBHs may be ejected from such environments by either dynamical recoil from binary--single scattering or gravitational-wave recoil following the merger of two black holes. 
We conduct Newtonian and post-Newtonian binary--single scattering experiments to study dynamical ejection in greater detail. 
We obtain fits to the probabilities for dynamical ejection, gravitational wave capture, and per-encounter hardening as a function of the binary mass ratio and hardness with respect to its environment. 
We borrow techniques from survival analysis (commonly used in studies of medicine, epidemiology, engineering, etc.) to develop a model to calculate the probability of IMBH binary ejection vs in-cluster merger. 
We confirm that the dynamical ejection probability strongly depends on both the mass ratio of the IMBH compared to other BHs in its environment and on the environment's velocity dispersion. 
We estimate that for a typical Milky Way globular cluster, IMBHs with mass $\lesssim10^3\,\msun$ are unlikely to be retained until the present. 
Our results also suggest that IMBH mergers with $q\lesssim0.2$ may be detectable at higher redshifts with future gravitational wave instruments such as the Einstein Telescope and Cosmic Explorer.

\end{abstract}

\section{Introduction} \label{sec:intro}
Intermediate-mass black holes (IMBHs) have masses between $10^2$ and $10^5\,\msun$ and remain elusive \citep[see][for a review]{imbhreview}.
Recently, there has been evidence for the presence of a black hole (BH) with mass $8,200-50,000\,\msun$ at the center of the Milky Way globular cluster (GC) Omega Centauri \citep{Haberle2024, Chen2025}.
On the other hand, IMBHs have been unambiguously detected through gravitational-wave (GW) transients, such as the recent event GW231123 with component masses of $137\,\msun$ and $103\,\msun$ \citep{gw231123}. 
At least one of these BHs falls within the ``upper-mass gap,'' the predicted range of roughly $40-120\,\msun$ where BHs should not form from isolated single-star evolution because pair-instability processes destroy the progenitor. 
However, several studies suggest that this region can be populated through consecutive stellar collisions \citep[e.g.,][]{DiCarlo19,Rizzuto,Kremer20,Banerjee_2020, EGP+2024} or hierarchical BH mergers \citep[e.g.,][]{MillerHamilton2002,McKernan2012,Rodriguez2018b,AntoniniGieles2019,GerosaBerti2019,FragioneLoeb2020,FragioneSilk2020}, among other mechanisms. 

Studies employing both direct $N$-body or Monte Carlo simulations of GCs show that, depending on the assumed initial conditions, IMBHs can be formed from the direct collapse of very massive stars (VMSs) produced from chains of stellar mergers in clusters, with the final mass of the of the VMSs very sensitive to initial density of the cluster, binary fraction of massive stars, and metallicity \citep[e.g.,][]{DiCarlo19,Rizzuto,Kremer20,Banerjee_2020, NW+2021, EGP+2021, EGP+2024, FujiiM+2024, Khurana2025, SharmaK+RodriguezCL2025, FROSTCLUST2, FROSTCLUST3}.
This channel has been theorized to produce stars with core masses below the pair-instability regime and oversized hydrogen envelopes, thus avoiding pair instability and instead collapsing into a BH \citep{Spera2019}. 
The mass of a BH born from the collapse of a VMS progenitor can then be much larger than is typically possible in isolated evolution channels, depending on the assumed fraction of the H envelope that falls back onto the BH \citep{CostaG+2022, BalloneA+2023}.

A key question is whether lower-mass IMBHs ($10^2-10^3\,\msun$) can be retained in their host environment from their formation at early times and continue to grow until the present day. 
For example, in the case of IMBHs forming from consecutive collisions in GCs, it has been shown that these objects are easily ejected from the clusters due to recoil kicks from both BH mergers and dynamical interactions \citep{EGP+2022}. 
However, if lower-mass IMBHs can be retained, it has been shown that they can grow to much larger masses through mergers with stellar-mass BHs and tidal disruption events \citep{RizzutoFP+2023, EGP+2025}.
If these objects can be retained in their host environments, they are an interesting source for a number of electromagnetic transient phenomena, including hypervelocity stars \citep[e.g.,][]{Brown2005, Boubert2018,Koposov2020}, tidal disruption events of stars \citep{Ramirez-RuizE+RosswogS2009, ChenJ+ShenR2018, KirogluF+2023}, and tidal disruptions of white dwarfs \citep{HaasR+2012, MacLeodM+2016}. 

Studies of initial conditions generally find that without an initially large number and density of massive stars, the formation of VMSs through stellar mergers is rare and IMBHs are formed almost entirely through hierarchical mergers. 
This is problematic because IMBHs which are formed hierarchically from the mergers of other BHs can have very large spin magnitudes, which greatly increases the magnitude of gravitational-wave recoil kicks \citep[e.g., ][]{RodriguezCL+2019}.
However, those formed from direct collapse IMBHs may be born with a small or negligible spin magnitude \citep{FullerJ+MaL2019}. 
A recent simulation with initial $N=10^7$, the largest to date, found that multiple-generation BHs can be formed efficiently in these clusters through hierarchical mergers, but this study, too, finds that the IMBHs produced are eventually ejected \citep{AM+2025}. 
The simulation conducted for this study had a wall-clock time in excess of $18$ months, making more detailed studies of the full parameter space in this range of $N$ and above impractical, even with Monte Carlo methods.

To circumvent these computational difficulties, a large number of studies develop and use semi-analytic methods for the population synthesis of dense stellar clusters to study the properties of BH subsystems \citep[e.g., ][]{AntoniniF+RasioF2016, AntoniniGieles2019,FragioneSilk2020,  FragioneG+2022, AtallahD+2023, ChattopadhyayD+2023, Kritos+2024}.  
Studies using such methods typically consider one or a population of BH binaries evolving along with a cluster background under H\'enon's Principle of balanced evolution \citep{BreenPG+HeggieDC2013}. 
Most studies find that only clusters with total mass $10^7\,\msun$ and above, e.g. the most massive GCs or nuclear star clusters, are efficient at retaining IMBHs and facilitating further growth due to their high escape speeds.

A common practice in studies employing semi-analytical methods is to evaluate the probability of ejection by assuming a fixed hardening rate per dynamical encounter and computing a critical semimajor axis $a_{\rm ej}$ below which a typical few-body interaction will eject the target binary.
However, we argue that, as binary ejection arises from the tail of the energy-change distribution, this assumption is worth revisiting explicitly using scattering experiments.
Furthermore, the scattering experiments conducted in the past---used to calibrate the hardening rates---tended to focus on either the limit where $m_1 = m_2 = m_3$ \citep[e.g., ][valid for typical LVK-type BBHs]{HutP+BahcallJN1983} or $m_1\approx m_2 \gg m_3$ to study binary SMBH hardening in galactic centers \citep[e.g., ][]{Quinlan+1996,SesanaA+2006,RasskazovA+2019}.
Some studies have been conducted looking at an intermediate range of mass ratios \citep[e.g.,][]{SS+PES1993, HeggieDC+1996}, though they predate the current interest in BH interactions and do not consider the effect of gravitational radiation.
More recently, \citet{RandoForastierB+2025} studied binary--single interactions of different mass ratios, but were primarily concerned with the calculation of BH mergers rates in more typical stellar environments, which do not produce IMBHs.

In this study, we investigate the impact of both dynamical recoil kicks from binary--single interactions and gravitational-wave recoil kicks on the retention and ejection of IMBHs in dense star clusters by considering a single IMBH-BH binary in a fixed background of single BHs with the same mass as the secondary.
Specifically, we seek to find an IMBH--BH mass ratio that is essentially immune to ejection.
For this purpose, we conduct a large suite of binary--single scattering experiments at intermediate mass ratios. 
In Section~\ref{sec:methods}, we describe our numerical methods. 
In Section~\ref{sec:results}, we show and describe our numerical results. 
In Section~\ref{sec:discussion}, we discuss the implications of our results on the detection of IMBH candidates as well as for current and future gravitational wave observations. 
Finally, in Section~\ref{sec:conclusion} we summarize our findings.

\section{Numerical Setup and Methods} 
\label{sec:methods}

\subsection{Setup for Individual Scattering Experiments}
In this study we use scattering code \texttt{Fewbody} to numerically evaluate the outcomes for different binary--single interactions \citep{fewbody}. 
As such, we will follow the notations outlined in \citet{fewbody}. 
In the context of binary--single interactions, the binary is denoted with subscript 0 (e.g., $a_0$, $e_0$) and the single is denoted $m_1$. 
When referring to individual members of the binary, the primary is denoted $m_{00}$ and the secondary $m_{01}$, such that the total mass of the binary $m_0 = m_{00} + m_{01}$. 

When considering the Newtonian, point-particle approximation, the outcome depends only on a few dimensionless parameters. 
The first is the ratio $b/a_0$ of the Keplerian hyperbolic impact parameter of the single to the binary semimajor axis. 
The next two are the mass ratio between the objects within the inner binary $q_{\rm in} = m_{01}/m_{00}$ and between the single and the primary $q_{\rm out} = m_1/m_{00}$ (equivalently, one could use the mass ratio between the single and the binary). 
The energy of the encounter is set by the ratio $\vinf/\vcrit$ where $\vinf$ is the Keplerian relative velocity at infinity and
\begin{equation}
    v_{\rm crit} = \sqrt{\frac{G}{\mu}\frac{m_{00} m_{01}}{a_0}}
\end{equation}
is the critical velocity for which the total energy in the encounter is 0; here $\mu = m_{0}m_{1}/(m_0 + m_1)$ is the reduced mass of the objects in the encounter. 
However, throughout later sections we will instead refer to the closely related ratio $\vorb/\vinf$ for easier physical interpretability, where $\vorb^2=\frac{Gm_0}{a_0}$ is the circular orbital speed of the binary.
When $m_{00}\gg m_{01}=m_1$, $\vcrit$ reduces to $\vorb$ exactly.
The relationship between $b$ and the actual distance of closest approach, $r_{\rm p}$, must take gravitational focusing into account:
\begin{equation}
    b = r_{\rm p}\sqrt{1+\frac{2G(m_0 + m_1)}{r_{\rm p} v_{\inf}^2}}.
    \label{eqn:b}
\end{equation}

For a single combination of the parameters $[q_{\rm in},q_{\rm out},\vorb/\vinf]$, we marginalize over the orbital orientation and phase, the eccentricity of the binary, and the impact parameter in order to create distributions of outcomes.
We sample the orientations and phases such that the encounters are isotropic. 
The binary eccentricity is sampled from a thermal distribution $p(e_0)\dd{e_0} = 2e_0\dd{e_0}$ \citep{JeansJ1919, Heggie1975}.
The impact parameter between the binary and the single is limited to a value $\bmax$, which is large enough that all outcomes of interest are captured and, for values $b>\bmax$, there are only flybys where the binary is weakly perturbed. 
Because of the geometry of the problem, larger impact parameters are more common than smaller ones. 
We sample the impact parameter $p(b)\dd{b} = 2\pi b \dd{b}$. For the value of $\bmax$, we use Eq.~(\ref{eqn:b}) with
\begin{equation}
    r_{\rm p, max} = 5\,a_0.
    \label{eqn:rpmax}
\end{equation}
We have verified that for all $q$ considered here our results would not change had we used larger values of $r_{\rm p, max}$.
In principle, in order to be fully self-consistent, it would be necessary to include an eccentricity dependence in this expression when considering individual scattering experiments. 
However, we elect to neglect this dependence for several reasons. 
First, it complicates comparisons with previous studies, which do not include any eccentricity dependence. 
Second, including this dependence would unnecessarily complicate the calculation of cross sections, since we marginalize over a thermal eccentricity distribution. 
We do not expect this to affect the final outcomes, as we are primarily concerned with large exchanges of energy between the binary and the single. 
Neglecting the eccentricity dependence in this way only increases the number of perturbative flybys for the relatively rare (near-)circular binaries. 

\subsection{Outcomes, Cross Sections, and Timescales}
\label{sec:outcomes}

The cross section for an outcome X (e.g., a binary energy change greater or less than some value) can be written as
\begin{equation}
    \crosssec{X} = \pi \bmax^2 \frac{N_\mathrm{X}}{N},
    \label{eqn:crosssec}
\end{equation}
where $N$ is the total number of experiments and $N_X$ is the number which result in an outcome $\mathrm{X}$.
In some cases, it is preferable to work with the branching ratios $\prob{X}=N_\mathrm{X}/N$, which can be interpreted as the empirical probability for the $\mathrm{X}$.
Since these branching ratios are computed numerically, there is some inherent uncertainty in the calculation. 
The first, from Poissonian counting statistics, is
\begin{equation}
    \Delta\crosssec{X} = \pi \bmax^2 \frac{\sqrt{N_\mathrm{X}}}{N}.
\end{equation}
The uncertainty $\Delta\prob{X}$ follows similarly.
The second, $\crosssec{inc}$, comes from encounters that are classified as incomplete due to reaching a wall-clock time limit of 1 hour. 
$\crosssec{inc}$ is calculated with Eq.~(\ref{eqn:crosssec}) above, so that the total uncertainty
\begin{equation}
    \Delta\crosssec{X,tot} = \Delta\crosssec{X} +\crosssec{inc},
    \label{eqn:error}
\end{equation}
but in practice $\prob{inc}\lesssim10^{-5}$ so that the Poissonian uncertainty usually dominates.

Given $\crosssec{X}$ for different outcomes, the relative rates of these outcomes can be estimated with $\rate{X} \approx n \crosssec{X} \vinf$. 
In the hard-binary limit ($\vorb\gg\vinf$) that we consider, the second term under the radical in Eq.~(\ref{eqn:b}) dominates, and this reduces to
\begin{equation}
    \rate{X} = \frac{2\pi n G (m_0 + m_1) \rpmax}{\vinf} \frac{N_X}{N}.
    \label{eq:rate}
\end{equation}
Under the hard-binary approximation, we can write the time between encounters, the inverse of the previous expression, as 
\begin{equation}
    \tenc = \frac{\vinf^3}{10\pi n G^2 m_{00}^2} \left(\frac{\vorb}{\vinf}\right)^2 \frac{1}{(1+q)(1+2q)}.
    \label{eqn:Tenc}
\end{equation}
While the outcomes of binary--single interactions are scale-free in isolation, the rates or timescales of interactions depend on the mass scale $m_{00}$. 
One factor of $m_{00}$ comes from enhanced gravitational focusing for larger masses. 
A second comes from the dependence on the physical size of the binary, which we have eliminated in favor of  $\vorb/\vinf$. 
The timescale for an outcome X and its associated uncertainty are then
\begin{equation}
\begin{split}
    T_{\rm X} &= \frac{N}{N_X}\tenc \\
    \Delta T_{\rm X} &= T_{\rm X} \frac{1}{\sqrt{N_X}}.
    \label{eqn:TX}
\end{split}
\end{equation}

\subsection{Escape from Clusters}

The previous discussion only considered binary--single encounters in isolation. 
However, escape can only be defined in relation to the host cluster. 
We make the approximation that all interactions happen in the center of the cluster core due to efficient mass segregation of the IMBH binary. 

For each encounter, we assume that the interaction velocity $\vinf$ is equal to the central velocity dispersion $\sigdisp$. 
In doing so, we set the relation between the energy scale of the cluster and the energy scale of the scattering interaction.
The relationship between $\vinf$ and the escape velocity $\vesc$ is determined by the depth and shape of the potential well. 

We use a Plummer profile \citep{Plummer1911}, characterized by a radial scale $a$, with a potential
\begin{equation*}
    \Phi(r) = -\frac{GM_{\rm tot}}{\sqrt{r^2 + a^2}}.
\end{equation*}
From the potential, the density 
\begin{equation}
    \rho(r) = \frac{3M_{\rm tot}}{4 \pi a^3}\left( 1 + \frac{r^2}{a^2}\right)^{-5/2},
\end{equation}
velocity dispersion 
\begin{equation}
    \sigdisp^2(r) = -\frac{\Phi(r)}{6},
\end{equation}
and escape velocity
\begin{equation}
    \vesc(r) = \sqrt{-2\Phi(r)} = \sqrt{12} \sigdisp(r)
\end{equation}
can be found \citep[e.g.,][]{BTtextbook}. 
In general, the velocities $\vinf$, $\sigdisp$, and $\vesc$ are related by numerical coefficients whose values are determined by the structure of the cluster potential.
In Section~\ref{sec:potential}, we will explore the impact of different choices for the potential.

To determine $\prob{ej}$, the fraction of objects which are ejected, we compare $\vesc$ with $\vkick$, the magnitude of the dynamical recoil from a binary--single interaction. 
In order to calculate $\vkick$, we first compute the magnitude of the new relative velocity from infinity
\begin{equation}
    \vinf'^2 = \frac{m_0m_1}{m'_0m'_1}\vinf^2 + \frac{2(m_0+m_1)E_0 \Delta}{m'_0m'_1}
    \label{eq:newvinf}
\end{equation}
where the primed masses denote the masses of the binary and the single following the interaction, as they may be different due to an exchange interaction, $E_0$ is the initial energy of the binary (e) and
\begin{equation}
    \Delta = \frac{E_0 - E'_0}{E_0}
\end{equation}
is the fractional change in the total energy of the binary following the interaction
\citep[we have the notation of Equations 4.1 and 4.3 of][]{SS+PES1993}. 
With this definition, $\Delta<0$ if a binary is hardened and $\Delta>0$ if it is softened.
Finally, $\vkick$ for the binary follows from Eq.~(\ref{eq:newvinf}) and momentum conservation:
\begin{equation}
    \vkick = \frac{m'_1}{m'_0+m'_1} \vinf'.
    \label{eq:vkick}
\end{equation}
A similar expression exists for the single.

\subsection{Triple Formation and Destruction}

In a binary--single encounter without dissipative forces, it is impossible for a stable triple to form as a consequence of time reversal symmetry \citep{chazy1929}.
However, in analogy to the formation of a tidal capture binary, temporarily bound triple systems can be formed during extremely long-lived resonance interactions \citep{Bailyn1989}.
For this reason, \texttt{Fewbody} reports a small but non-negligible triple formation cross section for large values of $\vorb/\vinf$. 
\texttt{Fewbody} determines the stability of a triple using the criterion \citep{MAstability}
\begin{equation}
\begin{split}
        \frac{a_{\rm out}}{a_{\rm in}} > \frac{3.3}{1-e_{\rm out}}\Biggl[ \frac{2}{3} \left( 1+\frac{m_{\rm out}}{m_{\rm bin}} \right) \\ \times \frac{1+e_{\rm out}}{(1-e_{\rm out})^{1/2}} \Biggr]^{2/5} (1- 0.3I/\pi)\, ,
\end{split}
\end{equation}
where in this notation $a_\mathrm{in}$ is the semimajor axis of the inner binary with mass $m_\mathrm{bin}$, $a_\mathrm{out}$ and $e_\mathrm{out}$ are the semimajor axis and eccentricity of the outer binary between $m_\mathrm{bin}$ and the tertiary $m_\mathrm{out}$, and $I$ is the mutual inclination between the two orbits. 
The ``stable'' triples reported by this criterion are, by construction of the scattering experiments, spurious in the sense that they are only metastable. 
This seeming contradiction is due to the construction of the stability criterion itself. 
This and later criteria \citep[e.g., ][]{LalandF+TraniAA2022, HayashiT+2022, HayashiT+2023} have been developed empirically through the use of extensive N-body modeling in order to determine the boundary between stability and instability. 
However, since all 3-body systems are chaotic, it is possible that the dissolution timescale for systems that fit these criteria on paper are much longer than the integration time used in these studies.

In principle, the production of hierarchical triples in our simulations represents a numerical uncertainty inherent to the classification system and should be included in the computational uncertainty budget of Eq.~(\ref{eqn:error}). 
However, in light of the discussion above, we could instead argue for a few different resolutions. 
On one hand, we can assume that these triples will be quickly ionized by the stellar environment and simply break them up, similar to the phenomenon highlighted by \citet{YBG+HP2021}. 
On the other hand, we can assume that eventually these triples will dissolve due to energy exchange. 
In this case, we can assume that the outer orbit will be ionized by extracting energy from the inner orbit. 
We check two different cases: first, we assume a negligible energy change $\vinf' = \vinf$. 
Second, we assume that the encounter terminates with a perfectly parabolic encounter such that $\vinf' = 0$, having softened the binary by the maximum possible amount. 
Regardless, we would not expect this to radically change $\crosssec{ej}$, which depends on the binary having hardened.
However, it is possible that this could affect the per--encounter mean hardening rate we compute in Section~\ref{sec: hardening}. 

For constructing our results, we choose the first option, since, when including post-Newtonian dissipation as we do in Section~\ref{sec: PN}, there is a small number of cases where significant dissipation occurs resulting in the formation of a truly stable triple system. 
These cannot be disentangled from the majority of cases where a likely misclassification occurred. 
We find that choosing between any of the first three cases results in no statistically significant difference in the results that follow.

\section{Results}
\label{sec:results}

In this section, we present results from numerical scattering experiments. 
We assume that a binary with primary mass $m_{00}$ and secondary mass $m_{01} = q m_{00}$ is in a fixed background of single masses $m_1 = m_{01}$.
We conduct scattering experiments over a grid with $q=[0.01,\, 0.02,\, 0.03,\, 0.06,\, 0.1,\, 0.2,\, 0.3,\, 0.6]$ and a range $\vinf/\vcrit=10^{-4}-10$ with 71 points evenly spaced logarithmically.
For each point in $(q,\vinf/\vcrit)$ space, we conduct $\mathcal{O}(10^4)$ scattering experiments.

Except where otherwise indicated, calculations for dimensionful quantities are done with fiducial values $m_{00} = 100\,\msun$, $\nfid$, and $\sigdisp \approx\vinf = 20 \, \rm{km/s}$. 
This density and dispersion is representative for young massive clusters based on the early-time properties of the models presented in \citet{cmccatalog, EGP+2022}. With this $\sigdisp$, $\vesc \approx 70 \, \rm{km/s}$ for a Plummer sphere.

\subsection{Key Physical Parameters}

\begin{figure*}
    \centering
    \includegraphics[width=0.92\textwidth]{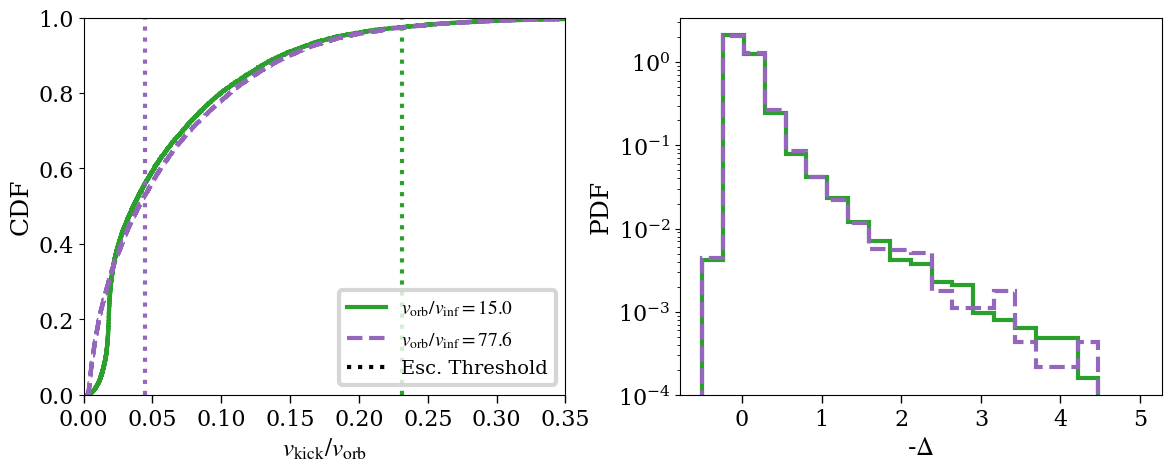} 
    \caption{(Left) Cumulative distributions of binary $\vkick/\vorb$ for different values of binary hardness $\vorb/\vinf$. 
    The vertical dashed lines shows the escape speed of the background cluster in the corresponding normalized units for the CDF of the same color. 
    Binaries with kick velocities to the right of this line are ejected. 
    (Right) Normalized distributions of the fractional energy change of the binary $\Delta$ as a result of the interaction for the same selected values of $\vorb/\vinf$. 
    Both sets of distributions show results from $q=0.6$ scattering experiments.
    In physical units, these selected values of $\vorb/\vinf$ correspond to $100\,\msun$--$60\,\msun$ binaries with semimajor axis $a_0=1.6\,\mathrm{au}$ (solid green) and $a_0=0.06\,\mathrm{au}$ (dashed purple) for $\vinf=20\,\mathrm{km/s}$.
    In the limit $\vcrit\gg\vinf$, the distribution of $\Delta$ converges to a single distribution. 
    Thus, in the normalized units used here, making the binary harder or softer corresponds to changing the boundary for escape. 
    For relatively soft binaries (green), the possibility of escape lies only in the tail of the recoil kick distribution, whereas for harder binaries (purple), escape may be possible in the bulk of the distribution.
    }
    \label{fig:hardness}
\end{figure*}

Here, we discuss how the distribution of the binary fractional energy change $\Delta$ and therefore of $\vkick$ is affected by some key physical parameters describing the interaction. 

The first key parameter is the binary hardness, which is characterized by the ratio $\vorb/\vinf$. 
This parameter characterizes the internal energy of the binary with respect to the kinetic energy of the single object. 
With our imposition $\vinf = \sigdisp$, the hardness parameter characterizes the energy scale of the binary with respect to the thermal energy of the cluster. 
In Figure~\ref{fig:hardness}, we show the influence of the hardness of the binary on the distributions $\vkick$ (left panel, cumulative density functions) and $-\Delta$ (right panel, probability density functions). 
Both sets of scattering experiments have the same mass ratio $q = 0.6$ but  different hardness parameters, with the solid green lines corresponding to the softer binary and the dashed purple lines to the harder binary. 
The dotted line in the top panel shows the value of the escape boundary for the cluster in the dimensionless units $\vkick/\vorb$ for each value of hardness. 
Objects to the right of this boundary can be said to escape. 
For convenience, we also state in the caption the semimajor axis of the binaries in each set of scattering experiments, assuming our fiducial $m_{00}$ and $\vinf$. 

For the case of a hard binary, the energy of the incoming single is negligible with respect to the energy of the binary. 
As a result, the PDFs shown in the bottom panel are roughly the same, with much of the difference occurring in the low-probability tail of the distribution. 
The $\vkick$ CDFs in the top panel are also roughly the same, with the high-velocity tail corresponding to the tail in the $\Delta$ distributions. 
Changing the value of the hardness parameter results in the escape boundary shifting to a higher or lower value. 
For the softer green distribution, the escape boundary is in the tail of the distribution, so only a few percent of the binaries here have escaped the cluster, whereas for the harder purple distribution, the boundary has moved into the bulk, so the escape fraction is approximately $45\%$.

\begin{figure}
    \includegraphics[width=0.46\textwidth]{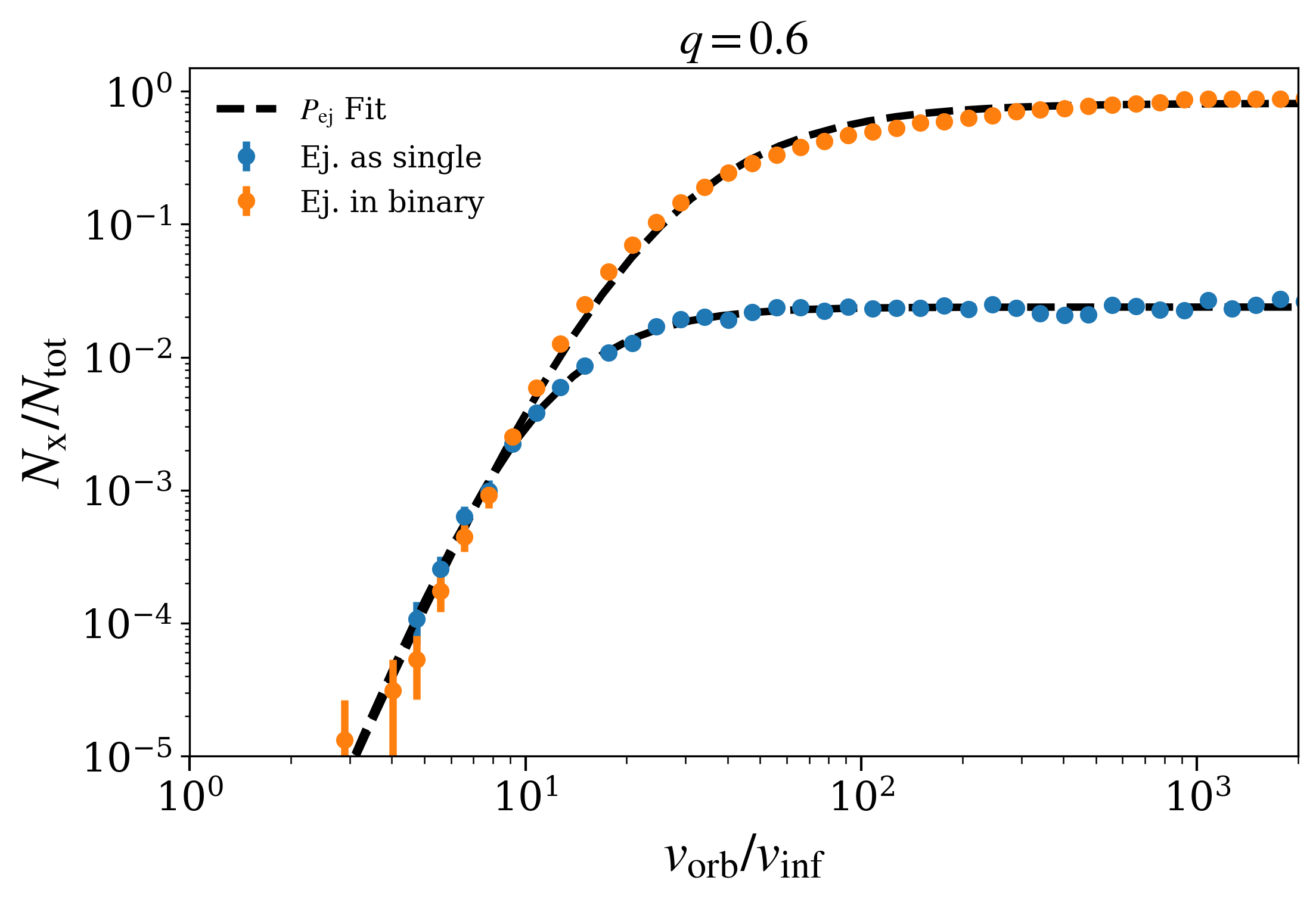}
    \caption{
    Branching ratios for ejection of the most massive BH in our $q=0.6$ interactions as a function of $\vorb/\vinf$.
    In most cases, the BH will be ejected as part of a binary (orange points), though in a minority of cases it may be ejected as a single object (blue points).
    We show the fit from Eq.~(\ref{eq:hillfit}) in black.
    The fit parameters for the fit to the binary ejection probability are provided in Tab.~\ref{tab:hillfit}.
    }
    \label{fig:branching_0.6}
\end{figure}

\begin{figure}
    \includegraphics[width=0.46\textwidth]{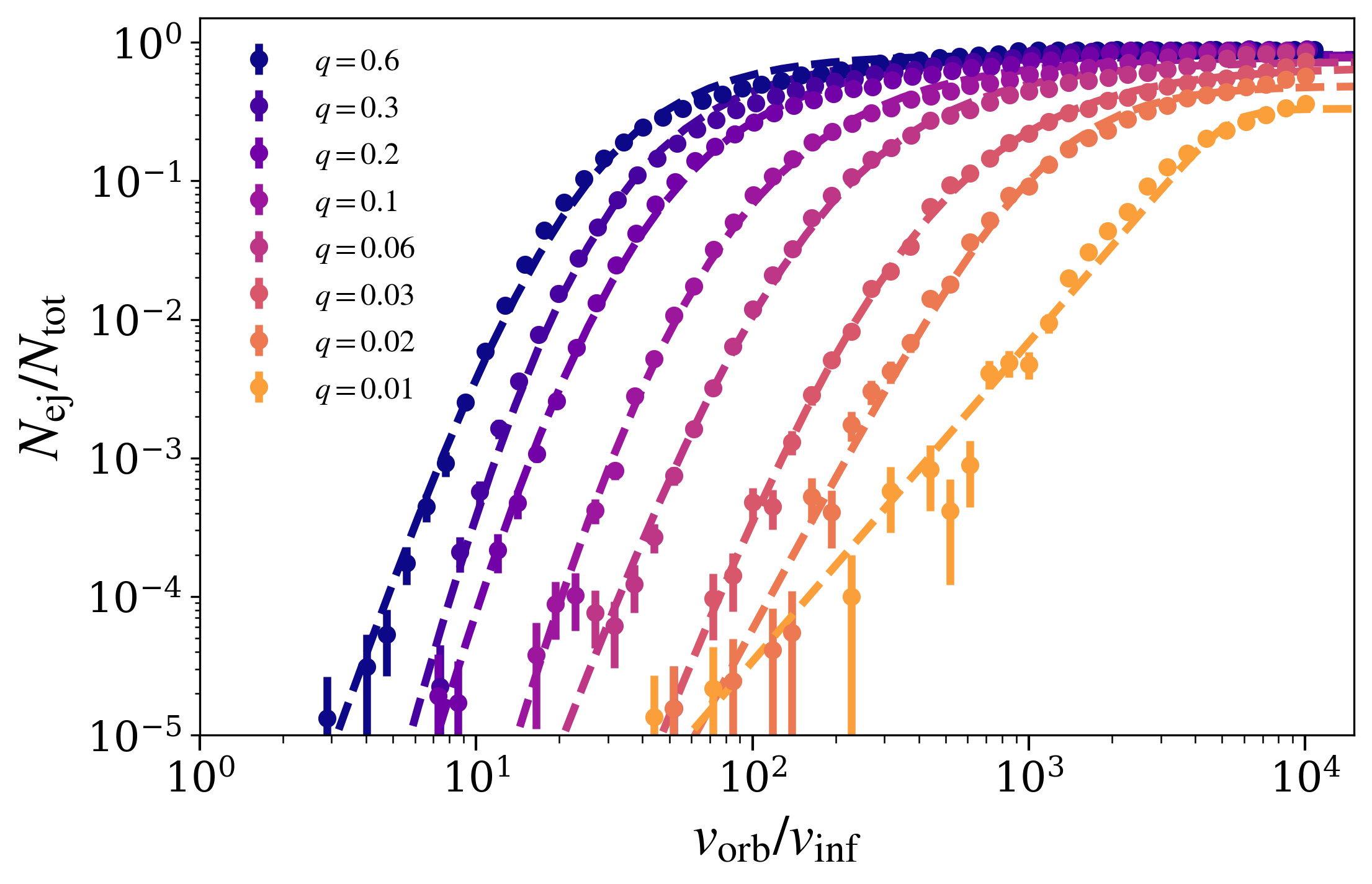}
    \caption{
    Branching ratios for ejection of the massive object in a binary for all simulated mass ratios (diamonds) with fits (dashed lines).
    Fit parameters are provided in Tab.~\ref{tab:hillfit}.
    }
    \label{fig:branching_q}
\end{figure}

In Figure~\ref{fig:branching_0.6}, we show the ejection branching ratio for $q=0.6$ as a function of $\vorb/\vinf$. 
In most cases, the most massive object is ejected as part of a binary (orange points), but in a minority of cases it is possible that it is ejected as a single object (blue points). 
The different data and fits correspond to the highest (left-most) to the lowest (right-most) mass ratios simulated.
In general, the probability for ejection of the most massive object as a single is negligible, so that in following sections, we define $\prob{ej}$ and the associated cross sections and timescales to refer only to cases where the most massive object is ejected as part of the final binary.
This preference arises from the usual heuristic that there is an energetic preference that the lightest object involved becomes the single following a strong interaction \citep[e.g.,][]{HeggieDC+1996}.
Furthermore, for constant $\Delta$, $\vkick$ is smaller when the massive object is ejected as a single by inspection of Eq.~(\ref{eq:newvinf}) and Eq.~(\ref{eq:vkick}) compared to when it is ejected as part of the binary.
This difference between the two branching ratios becomes more stark with lower values of $q$, which is already quite dramatic for the largest $q$ tested.

In Figure~\ref{fig:branching_q}, we show the branching ratio for ejection in the binary for all mass ratios as a function of $\vorb/\vinf$.
In both figures, we also present a fit of the form 
\begin{equation}
    \label{eq:hillfit}
    \prob{ej} = A\left[1 + \left(\frac{w}{\tilde{v}}\right)^{n} \right]^{-p/n}
\end{equation}
where $\tilde{v} = \vorb/\vinf$.
Fit parameters\footnote{While the fits provided are adequate for the purpose of this study, we caution against quoting the exact numerical values for theoretical work. Because the low velocity tail is undersampled, the parameters are somewhat degenerate.} are given for all mass ratios in Table~\ref{tab:hillfit}.
The different parameters have the following interpretation.
At low $\tilde{v}$, the function behaves as $\tilde{v}^{-p}$, whereas at high $\tilde{v}$ it saturates to the amplitude $A$.
Then, $w$ is approximately the location where the transition occurs, and $n$ controls the sharpness of the transition.
The velocity at which it becomes probable for the binary to be ejected (say, $\prob{ej}\gtrsim10\%$) moves further to the right with decreasing $q$, as expected.

\begin{table}
    \centering
    \begin{tabular} {c | c c c c }
    \hline
    $q$ & $A$ & $w$ & $n$ & $p$ \\ \hline
    $0.6$ & $0.81$ & $21$ & $1.5$ & $5.8$ \\
    $0.3$ & $0.78$ & $20$ & $1.4$ & $8.3$ \\
    $0.2$ & $0.81$ & $26$ & $1.2$ & $7.9$ \\
    $0.1$ & $0.79$ & $44$ & $1.1$ & $8.3$ \\
    $0.06$ & $0.73$ & $1.4\times10^2$ & $1.2$ & $5.6$ \\
    $0.03$ & $0.67$ & $3.5\times10^2$ & $1.2$ & $5.3$ \\
    $0.02$ & $0.49$ & $1.2\times10^3$ & $2.0$ & $3.7$ \\
    $0.01$ & $0.33$ & $5.4\times10^3$ & $6.8$ & $2.3$ \\ \hline
    \end{tabular}

    \caption{Fit parameters for Eq.~(\ref{eq:hillfit}). }
    \label{tab:hillfit}
\end{table}

One point of interest is the importance of resonant interactions, which are named in analogy to particle physics.
Resonant interactions refer to those interactions which are long-lived compared to the typical crossing time of the interaction $\tint = a_0/\vinf$.
Such interactions are long-lived as a result of the formation of temporary hierarchical triples punctuated by democratic phases where there is no clear hierarchy.
However, there is no clear threshold in time separating the non-resonant and resonant interactions.
For this reason, an alternative definition is sometimes used based on counting the number of local minima of the sum of the squared pairwise distances between the objects in the interaction \citep{McMillanSLW+HutP1996}, but the threshold used in counting the minima with this definition is also somewhat arbitrary.
In any case, in the near equal-mass limit, there is an expectation for a clear divergence in the behavior of the interactions which are long-lived and those which are not \citep{HutP+BahcallJN1983, ManwadkarV+2020, TraniAA+2024}.

\begin{figure*}
    \centering
    \includegraphics[width=0.46\textwidth]{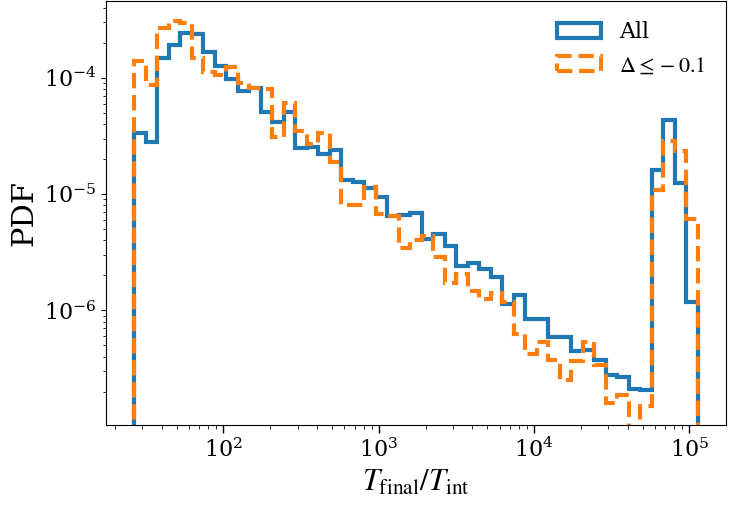}
    \includegraphics[width=0.46\textwidth]{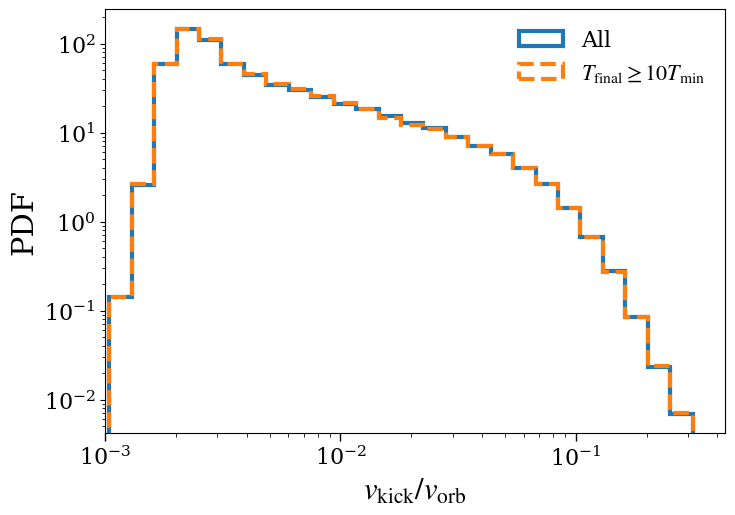}
    \caption{Distributions of $\tfinal/\tint$ (left) and normalized $\vkick$ (right) for all interactions (solid blue) and subsets of the interactions (dashed orange) with $q=0.2$ and $\vorb/\vinf \approx73$. 
    In the left panel, the subset includes only interactions for which the change in energy $\Delta\leq-0.1$, whereas in the right panel, it includes only interactions for which $\tfinal\geq10\,\tmin$.
    }
    %
    \label{fig:res}
\end{figure*}

In Figure~\ref{fig:res}, we show in the left panel the distribution of simulation times normalized by $\tint$, and in the right panel the distribution of normalized $\vkick$ for scattering experiments with $q=0.2$ and $\vorb/\vinf \approx73$. 
In either panel, the solid blue histogram corresponds to the distribution containing all interactions, whereas the dashed orange histograms only includes a subset of interactions for which $\Delta\leq-0.1$ (left) or simulation times $\tfinal \geq 10\,\tmin$ (right), where $\tmin$ is defined as the simulation time for a pure two-body hyperbolic encounter, which depends on the tidal perturbation $\delta$ parameter set in \texttt{Fewbody}. 
The results are not sensitive to the exact value of the threshold $10\,\tmin\approx100\,\tint$, which was simply chosen to safely exclude all the shortest-lived interactions.
The spike in the left panel for large values of $\tfinal$ corresponds to a typical orbital period for wide triples formed from an initially-softening encounter.
The results presented here are only for one set of simulations but the qualitative features are generic to all the simulations we ran.

However, this does not seem to be the case here, as shown by the equality in the two sets of histograms.
A detailed investigation for why this is the case is outside the scope of this work, though we offer some speculation.
We can understand this result intuitively in the following way: as these resonant interactions involve a weak perturbation resulting in a very wide triple, subsequent outer pericenter passes will likely result in similarly weak perturbations.
This eventually diffuses into the a configuration where the outer object is ionized.
For a terminal close encounter to result in some large kick, some degree of luck is still involved.
Incidentally, the described process is exactly equivalent to the conditions required to generate a L\'evy flight, which produces the power-law distribution seen in the right panel of Fig.~\ref{fig:res} \citep{ManwadkarV+2020}.

\subsection{Dependence on Background Cluster Potential}
\label{sec:potential}

\begin{figure}
    \centering
    \includegraphics[width=0.46\textwidth]{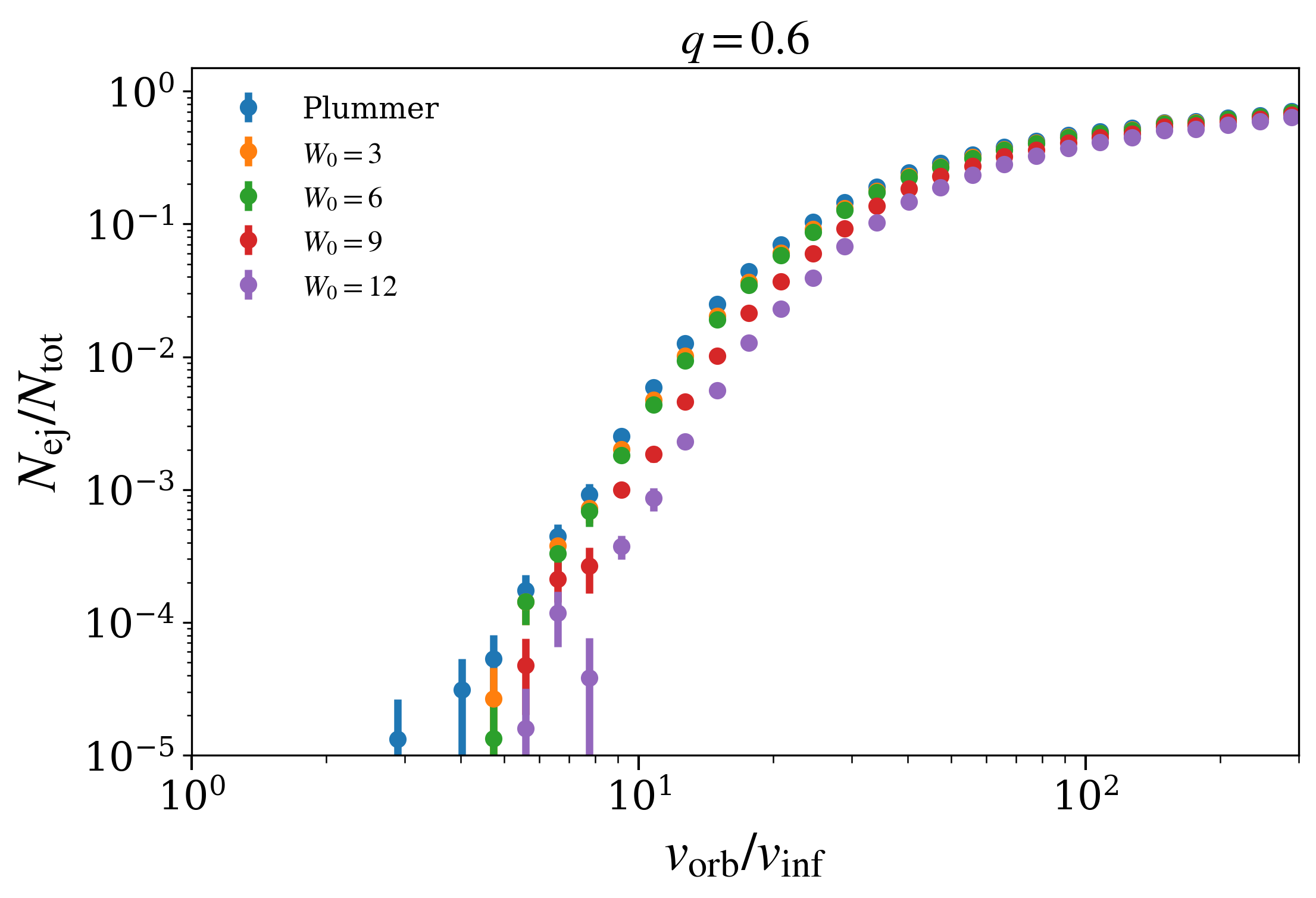}
    \includegraphics[width=0.46\textwidth]{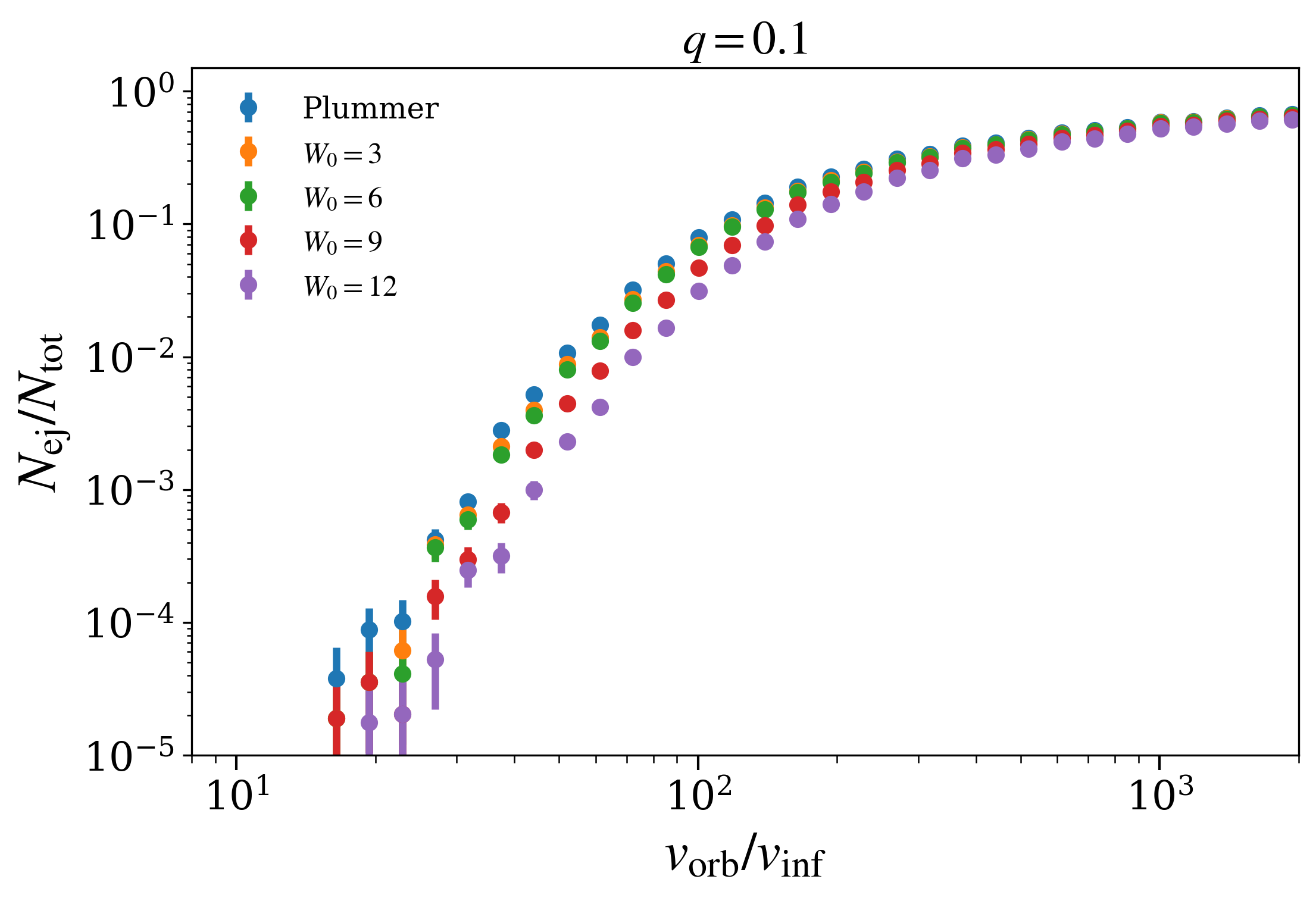}
    \caption{
    Branching ratios for mass ratios $q = 0.6$ (top) and $q = 0.1$ (bottom) with different assumptions of the background cluster potential as a function of binary hardness. 
    The Plummer model is shown in blue. The King models are shown in green, orange, red and purple with their associated $W_0=3,\,6,\,9,\,12$ respectively. 
    The range of $W_0$ values chosen correspond to a representative range of measured concentration parameters $c=\log(r_t/r_0)$ reported by the Harris catalog of Milky Way GCs \citep{HarrisCatalog}.}
    \label{fig:plummervking}
\end{figure}

As stated previously, the results shown thus far have assumed a Plummer potential for the background cluster.
While the Plummer profile is valued for its simplicity, it is unfortunately unrealistic. 
For instance, the Plummer model does not have a tidal cutoff radius, unlike observed clusters. 
A more realistic model is the family of King profiles \citep{King1962}, which is commonly used to characterize Milky Way GCs \citep[e.g., ][]{HarrisCatalog}. 
In Appendix~\ref{sec:king}, we derive the escape speed and velocity dispersion of a King profile as a function of the parameter $W_0$, which characterizes the central depth of the cluster's potential well.
We compare the probability of escape between these different models for the potential.

In Figure~\ref{fig:plummervking}, we show escape branching ratios as a function of hardness for both $q=0.6$ and $q=0.1$.
In both panels, it is apparent using a Plummer profile results in the least conservative (i.e. largest) estimate for $\prob{ej}$. 
The difference between the least conservative estimate and the most conservative estimate (King profile with $W_0=12$) is smallest for the hard binaries and less asymmetric mass ratios, as these conditions lead to more interactions which result in recoil kicks with larger magnitude than the escape speed. 
The difference is more apparent for softer binaries. 
In the most extreme cases, a branching ratio of 0 is reported for the $W_0=12$ criterion whereas other criteria report a non-zero branching ratio. 
However, as the reported cross section is already minimal in these cases, the difference does not greatly impact the reported results. 
Additionally, the differences in the cross section between the Plummer, $W_0=3$, and $W_0=6$ profiles are order unity across all values of $\vorb/\vinf$.
However, the shape of the curves, specifically the slope of the power-law, do not seem to change, so that the value for $w$ reported in Tab.~\ref{tab:hillfit} is different only by a constant factor.

\subsection{Hardening}
\label{sec: hardening}

\begin{figure}
    \centering
    \includegraphics[width=0.46\textwidth]{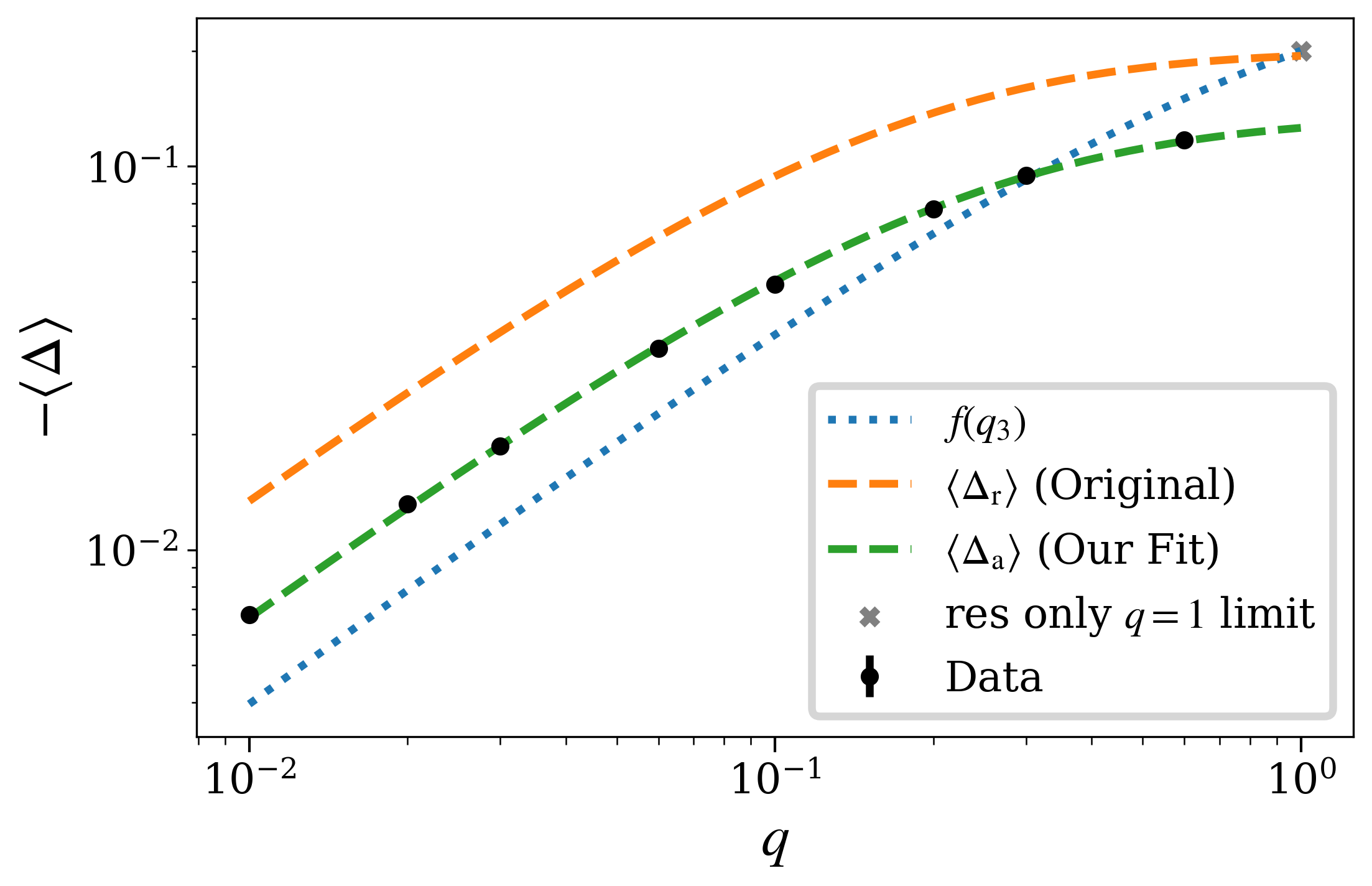}
    \caption{Average per-encounter change in binding energy $-\Deltaave$ as a function of $q$ in the hard-binary limit. Fits (black circles) are done from the point where $\Deltaave$ approaches a constant value, with error bars smaller than the points. For comparison, we show functions from the literature, namely Eq.~(\ref{eq:fq3}) from \citet{ChattopadhyayD+2023} (dotted blue), Eq.~(\ref{eq:deltaaveres}) from \citet{RandoForastierB+2025} (dashed green), and the value from \citet{HutP+BahcallJN1983} which was found for resonant interactions in the equal-mass case (grey x). We also refit Eq.~(\ref{eq:deltaaveres}) using our data for all interactions (dashed orange).}
    \label{fig:newtonharden}
\end{figure}

The ejection timescale implied by Eq.~(\ref{eqn:TX}) is not quite correct, since it implicitly assumes that the binary stays at a fixed $a_0$ (or equivalently, a fixed $\vorb/\vinf$). 
However, the encounters that do not eject the binary will still on average harden the target binary by some small but fixed amount. 

We find that in the hard-binary limit, the per-encounter average hardening $\Deltaave$ is constant for fixed $q$, consistent with previous studies.
This follows from the fact that the kinetic energy of the single is negligible compared to the internal energy of the binary.
For each $q$, we perform the straight-line fits in the hard-binary regime and present the results as the black circles in Figure~\ref{fig:newtonharden}.
The uncertainty in the fit is of the order $\sigma\sim10^{-4}$, such that the error bars are smaller than the points.
We also show the usual assumption for the $q=1$ case, that in the equal-mass limit $\Deltaave=-0.2$ for resonant encounters (gray x) \citep{HutP+BahcallJN1983}.

For the sake of comparison, we show two different functions from the literature.
First, \citet{ChattopadhyayD+2023}, for use in their semi-analytic code, propose a function of the form
\begin{equation}
    f(q_3) = - 0.4\, \frac{q_3}{1-q_3}
    \label{eq:fq3}
\end{equation}
with $q_3 = m_1/(m_0+m_1) = q/(1+2\,q)$, where the prefactor was chosen to converge to the result found in scattering experiments conducted in the equal mass case (gray x).
The function does not reflect the functional form of the data, though it is still correct within a factor of a few.

Second, we compare to the function found in the study by \citet{RandoForastierB+2025},
\begin{equation}
    \langle\Delta_{\rm r}\rangle = -\Delta_0 \left[1-\exp\left(-A\frac{m_1}{m_0}\right) \right]
    \label{eq:deltaaveres}
\end{equation}
with $\Delta_0 = 0.2$ (to match \citet{HutP+BahcallJN1983}) and $A=7$, where the subscript indicates that this fit was done for \textit{only resonant} interactions. 
At some level, the comparison is not quite valid, as we perform the average over \textit{all} interactions, whereas theirs is for only the resonant interactions.
However, our discussion of Fig.~\ref{fig:res} above shows that this should not make a significant difference.
The apparent disagreement is also not a concern, since \citet{RandoForastierB+2025} fit this function for the condition $m_1\leq m_0$, whereas our choice of parameters is much more restricted. 
Looking at their Fig. 1, we see a similar range of disagreement.
Nevertheless, the functional form looks similar to our averages over all data, motivating us to fit our own function $\langle\Delta_{\rm a}\rangle$, where the subscript indicates that this fit is done for \textit{all} interactions.
We find 
\begin{equation}
    \langle\Delta_{\rm a}\rangle = -0.14 \left[1-\exp\left(-5\frac{m_1}{m_0}\right) \right].
    \label{eq:deltaave}
\end{equation}

\subsection{Relativistic Effects}
\label{sec: PN}

\begin{figure}
    \centering
    \includegraphics[width=0.46\textwidth]{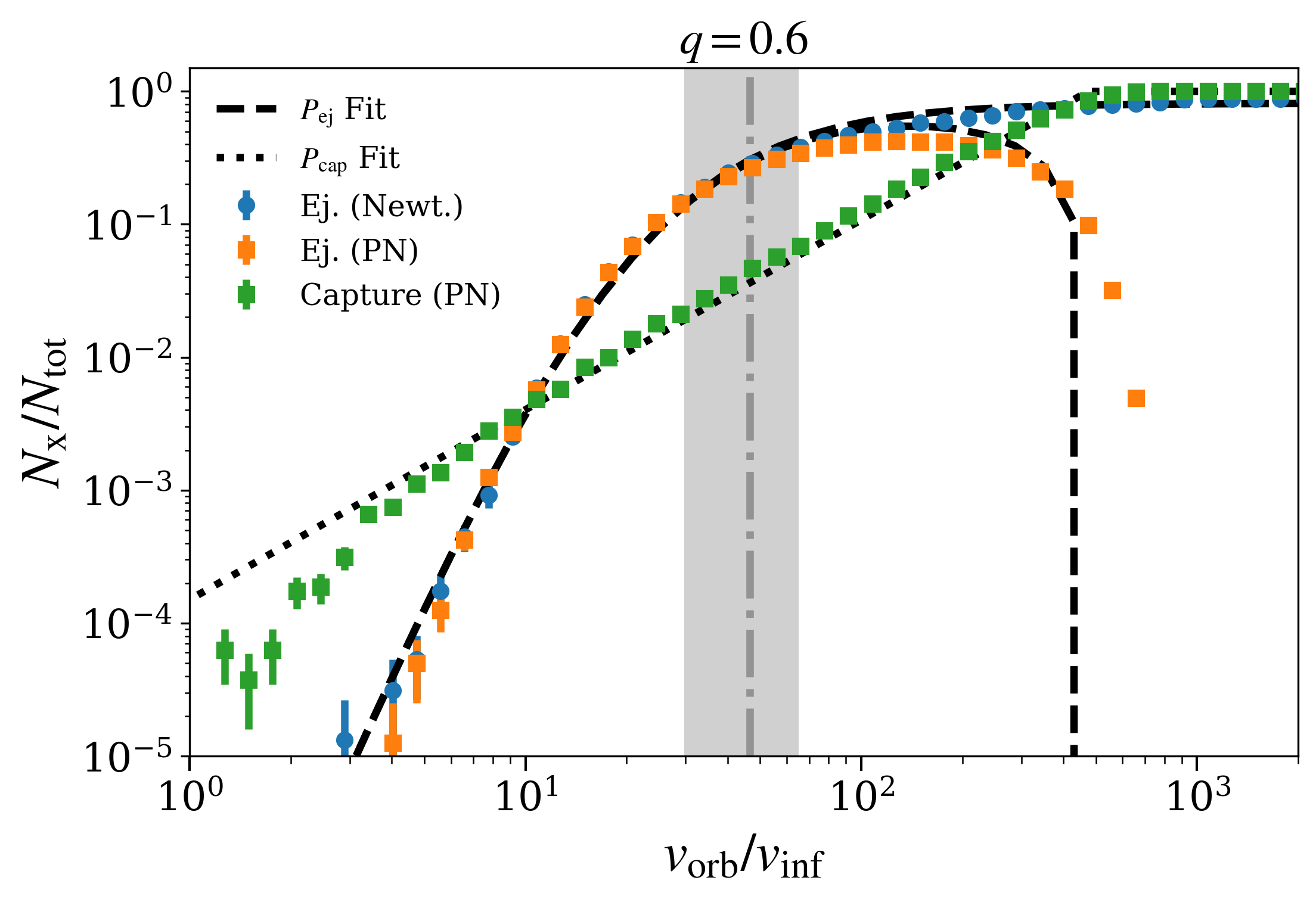}
    \includegraphics[width=0.46\textwidth]{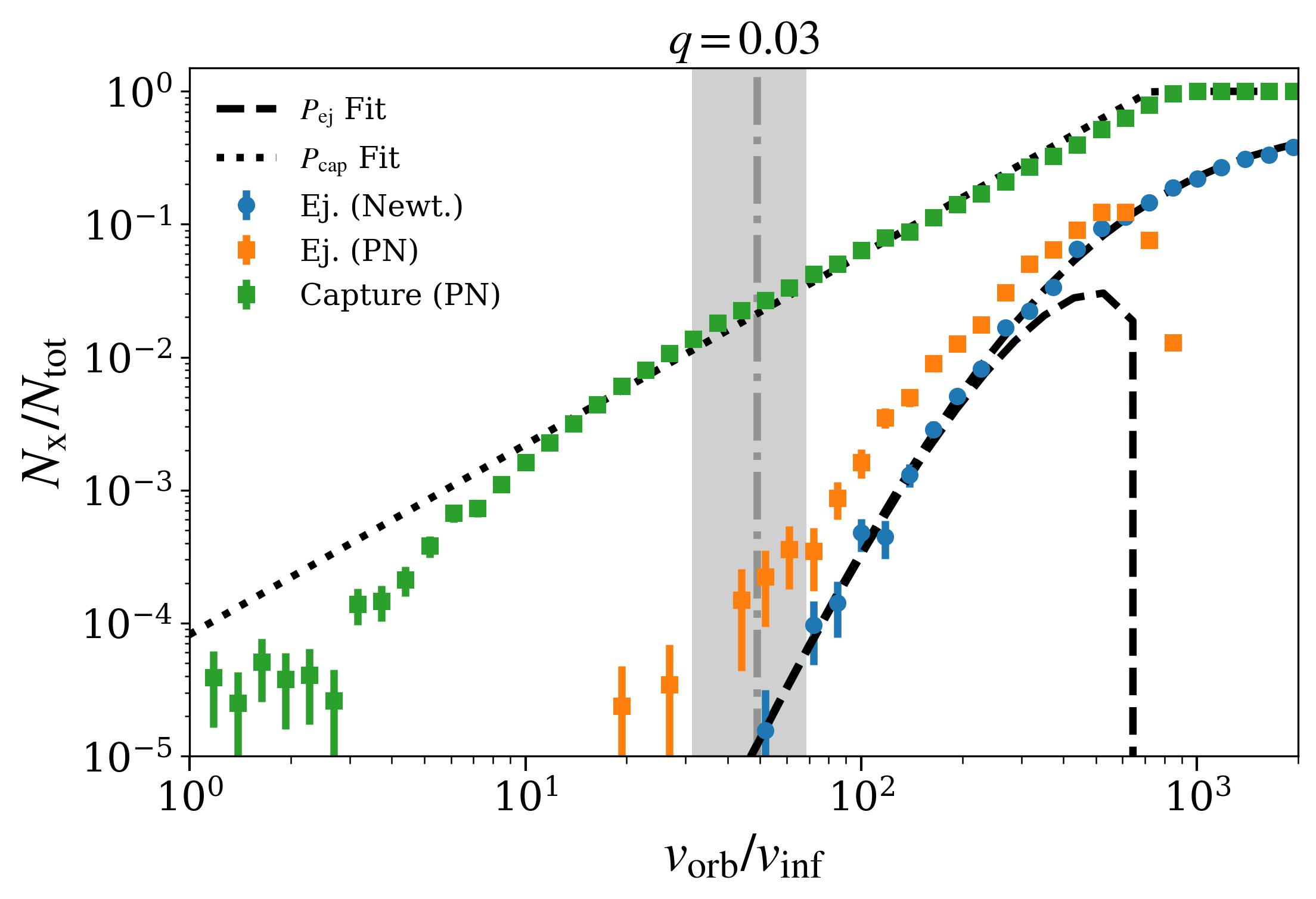}
    \caption{
    Branching ratios for $q=0.6$ (top) and $q=0.03$ (bottom) scattering experiments including PN terms and $\vinf = 20\, \rm{km/s}$. 
    The branching ratios for ejection and merger are shown with orange and green squares, respectively.
    The Newtonian data from Fig.~\ref{fig:branching_q} are replotted (blue circles) for comparison. 
    We also show the fit to $\prob{cap}$ with Eq.~(\ref{eq:p_cap}) with $B=2.61$ (dotted black lines).
    We attempt to match the PN ejections by combining a simple cutoff with the fit to the Newtonian data (dashed black lines), though this fails for $q\lesssim0.03$.
    Finally, the grey shaded areas correspond to the limit $\tenc/\tgw = 1$ (Eq.~\ref{eq:tenc_limit}) for $e_0 = 0 -0.9$, with the dashdotted line corresponding to $e_0=0.7$, roughly the median of the thermal distribution.
    To compute these timescales we assume $m_{00} = 100\,\msun$ and $\nfid$.
    }
    \label{fig:pn_scattering}
\end{figure}

At 2.5 Post-Newtonian (PN) order, the binary systems experience gravitational-wave radiation reaction, causing the binary to slowly shrink until the two objects merge. The inspiral time of a binary is given by the equation \citep{Peters1964}:

\begin{equation*}
    \tgw = \frac{12}{19} \frac{c_0^4}{\beta} \int_0^{e_0} \frac{e^{29/19}[1+(121/304)e^2]^{1181/2299}}{(1-e^2)^{3/2}} \dd e\,.
\end{equation*}

The constants $c_0$ and $\beta$ are
\begin{equation*}
c_0 \equiv a_0 e_0^{-12/19} (1-e_0^2) \left[ 1 + \frac{121}{304} e_0^2 \right]^{-870/2299}
\end{equation*}
and 
\begin{equation*}
\beta \equiv \frac{64}{5} \frac{G^3 m_{00} m_{01} m_0}{c^5}\,. 
\end{equation*}
In terms of $\vorb$ and $q$, this may be rewritten as 
\begin{equation}
    \tgw = \frac{15}{304} G m_{00} c^5 \vorb^{-8} \frac{(1+q)^3}{q} F(e_0)
\label{eqn:tpeter}
\end{equation}
where $F(e_0)$ is a function of the initial eccentricity $e_0$:
\begin{equation*}
\begin{split}
    F(e_0) = \left( e_0^{-12/19} (1 - e_0^2) \left[ 1 + \frac{121}{304} e_0^2 \right]^{-\frac{870}{2299}} \right)^4 \\ \times \int_0^{e_0} \frac{e^{29/19}[1+(121/304)e^2]^{1181/2299}}{(1-e^2)^{3/2}} \dd e\,
\end{split}
\end{equation*}
for $0 < e_0 < 1$ and $F(0) = 19/48$.

We conduct additional scattering experiments with $2.5$PN terms for all values $(q,v)$ for comparison.
We exclude the conservative $1$PN and $2$PN terms as they are not expected to affect on the cross sections of interest when secular effects are unimportant \citep{Rodriguez2018b}.
In order to detect mergers between objects, we assign each object an effective radius $10M$ in gravitational units.

There are two effects relevant for this study.     
First, similar to other finite-size effects, gravitational-wave radiation imposes an effective radius for each of the objects, such that mergers between objects can occur if they pass close enough for an impulsive, large emission of gravitational radiation.
During resonant encounters, repeated interactions between the three objects allow the inner binary to ergodically explore its allowed phase space.
Some of the possible configurations result in high eccentricities leading to a prompt GW-induced inspiral, resulting in mergers even when the typical velocities involved are non-relativistic \citep{SamsingJ+2014}.
From this, we can write the probability of a merger during a binary--single encounter, typically referred to as a capture, as \citep{SamsingJ2018, RandoForastierB+2025}
\begin{equation}
\begin{split}
    \prob{cap} &\simeq B\langle N_{\rm IMS} \rangle \frac{2r_{\rm GW}}{a_0} \\
    &\simeq5.36 B \langle N_{\rm IMS} \rangle q^{2/7} (1+q)^{-4/7} \left(\frac{\vorb}{c}\right)^{10/7}
\end{split}
\label{eq:p_cap}
\end{equation}
where $\langle N_{\rm IMS} \rangle$ is the number of intermediate hierarchical triple states in a resonant encounter and $r_{\rm GW}$ is the pericenter distance at which a single passage radiates energy equal to $E_0$ \citep{HansenRO1972}.
We also include a constant $B$ to fit to the data.
In principle, $\langle N_{\rm IMS} \rangle$ is a function of the mass ratios.
In the equal-mass case, $\langle N_{\rm IMS} \rangle\simeq20$ \citep{SamsingJ2018}, but in the case $m_{00}>m_{01}=m_1$, $\langle N_{\rm IMS} \rangle\simeq15$ for $0.1\lesssim q \lesssim 0.6$ \citep{RandoForastierB+2025}.
When $m_{01}\neq m_1$, $\langle N_{\rm IMS} \rangle$ can be much lower, such that our results are somewhat overpredicting the probability of captures \citep[for additional discussion see][]{RandoForastierB+2025}.

Second, we compare $\tgw$ to $\tenc$. If the quantity
\begin{equation}
\begin{split}
    \frac{\tenc}{\tgw} = \frac{304}{150\pi} \frac{\vinf^6}{n G^3 m_{00}^3} \left(\frac{\vinf}{c}\right)^5 \left(\frac{\vorb}{ \vinf}\right)^{10} \\ \times \frac{q}{(1+q)^4 (1+ 2q)} \frac{1}{F(e_0)}
\end{split}
\label{eq:tenc_limit}
\end{equation}
is larger than 1, then, on average, the binary will merge before experiencing its next encounter.

In Figure~\ref{fig:pn_scattering}, we show branching ratios for ejection (orange squares) and merger (green squares) for scattering experiments including $2.5$PN terms with $\vinf=20\,\rm{km/s}$ for $q=0.6$ (top panel) and $q=0.03$ (bottom panel). 
The Newtonian ejection branching ratios previously shown in Fig.~\ref{fig:branching_q} are shown for comparison (blue circles). 
We show with dashed black lines fits to both the PN data (described below) and the Newtonian data (from Tab.~\ref{tab:hillfit}).
We also show the fit to Eq.~(\ref{eq:p_cap}).
Finally, we show a grey shaded region where $\tenc/\tgw=1$ for $e_0=0-0.9$, with the dash-dotted line corresponding to $e_0=0.7$.
To compute $\tenc$, we assume $m_{00}=300\,\msun$ and $\nfid$.

We find that the probability for merger during the encounter follows the expected $P\propto v^{10/7}$ scaling for all mass ratios tested, but only for sufficiently hard velocities ($\vorb/\vinf\gtrsim3$).
However, we find that $B=2.61$ is required for Eq.~(\ref{eq:p_cap}) to fit the amplitudes.
There is still some $q$ dependence that is not captured by this constant factor, but with this value of $B$, Eq.~(\ref{eq:p_cap}) is correct to within $10\%$ for the range of $q$ simulated.
Note that \citet{RandoForastierB+2025} found that $\langle N_{\rm IMS} \rangle$, and therefore $\prob{merge}$, is maximized when $m_{01} = m_1$.
As a result, when considering more realistic BH mass functions where the two small masses close but unequal, $\prob{cap}$ will be lower by a factor of at most a few, or lower by about an order of magnitude when they are unequal.

Comparing the Newtonian and PN ejection probabilities, we see that in the top panel, the probability stays roughly the same until a divergence at roughly $\tilde{v}\approx60$, followed by a sudden drop-off when $\prob{cap}$ begins to approach unity.
We find that the PN $\prob{ej}$ is well described by taking the fit to the Newtonian data (upper dashed black lines) as reported in Tab.~\ref{tab:hillfit} and including a coefficient $1-\prob{cap}$ (lower dashed black lines).
Indeed, in the top panel, the discrepancy begins to be apparent by eye when $\prob{cap}\approx0.1$.
However, this fix is slightly inadequate for $\prob{cap}\simeq1$.
This is because $\prob{cap}$ is appropriate for the high-eccentricity limit, but not when $\tgw(e_0=0)$ is comparable to the interaction length. 
However, for $q\lesssim0.03$, we find that the ejection probability actually increases for constant $\vorb/\vinf$ when compared to the Newtonian case.
This discrepancy arises because, for small $q$, ejection is only possible when the binary is initially very hard. 
This also increases the amount of hardening via GW radiation during long resonant encounters, such that the final binary is even harder than the initial state, resulting in a larger $\vkick$.
However, the binary will generally merge between successive encounters prior to the point at which PN terms significantly impact the ejection probability.
Thus, we can treat the two branching ratios $\prob{ej}$ and $\prob{cap}$ as independent.

\subsection{Survival Analysis for Ejected Binaries}
\begin{figure}
    \centering
    \includegraphics[width=0.46\textwidth]{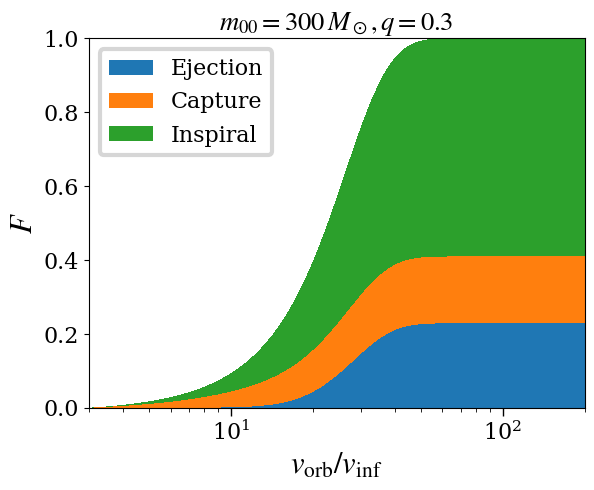}
    \caption{
    Stacked cumulative incidence functions for $m_{00} = 300\,\msun$, $q=0.3$, $\vinf = 20\,{\rm km/s}$, and $\nfid$.
    Here, the velocity acts as a time-like variable for a typical cumulative density function.
    The relative height of each colored region corresponds to the probability of the given outcome happening before the given binary hardness, such that when the three functions sum to $1$ all binaries have either merged or have been dynamically ejected.
    The different colors correspond to ejection (blue, $23\%$), capture (orange, $18\%$), and inspiral (green, $59\%$), respectively.
    }
    \label{fig:cumulative_ex}
\end{figure}

\begin{figure*}
    \centering
    \includegraphics[width=0.92\textwidth]{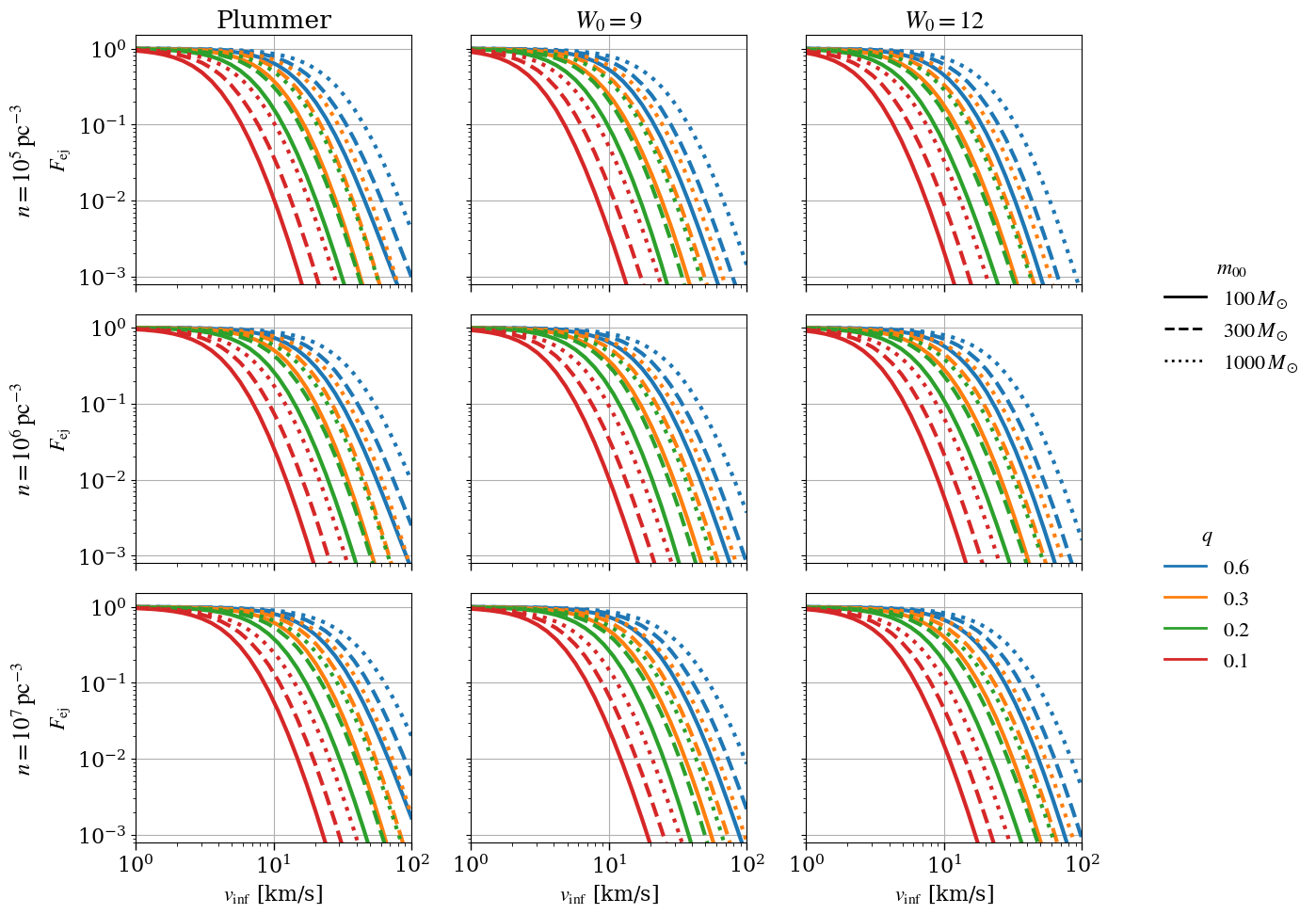}
    \caption{
    Maximum value of $F_{\rm ej}$ as a function of $\vinf$ for different combinations of parameters.
    Each column represents a different assumed potential, with the most optimistic Plummer potential on the left and the most conservative $W_0=12$ King profile on the right.
    The rows increase in number density from top to bottom from $10^5$ to $10^7\,{\rm pc}^{-3}$/
    Blue, orange, green and red lines correspond to $q=0.6, 0.3, 0.2,$ and $0.1$, respectively.
    Solid, dashed, and dotted lines correspond to $m_{00} = 100\,\msun,\,300\,\msun$, and $1000\,\msun$, respectively.
    }
    \label{fig:f_ej}
\end{figure*}

Survival analysis is a set of statistical techniques used to analyze the time until events of interest and is commonly used in fields such as epidemiology, medicine, and engineering.
We provide a brief introduction to the subject in Appendix~\ref{sec:survivalanalysis} for readers who are unfamiliar with these techniques.

In order to evaluate the probability of a given outcome occurring, it is not sufficient to compare the timescales or cross sections calculated from the above scattering experiments as there are three separate outcomes that scale differently with $\vorb/\vinf$.
Furthermore, the three outcomes are mutually exclusive, which in the language of survival analysis is described as {\it censoring}; if a binary merges between successive encounters, then the other two outcomes become forbidden by definition.
This, among other considerations, motivates the use of semi-analytic codes to calculate outcome probabilities.
However, under suitable simplifications, this problem can be cast in fully analytic terms using the fits from previous sections.

In this section, we use the numerical results of the previous section, as well as the methods of competing risk analysis described in Appendix~\ref{sec:survivalanalysis}, to construct hazard functions and determine the distribution of outcomes for IMBH binaries with different parameters.
We can construct the cause-specific hazard function for cause $C$ with a given set of parameters $\theta = (m_{00}, q, \vinf, n)$:
\begin{equation}
    h_C(\vorb|\theta) = \rate{C} (\vorb' | \theta) \left|\frac{\dd E_0}{\dd t}\right|^{-1} \frac{\dd E_0}{\dd \vorb'}.
\end{equation}
As $\vorb$ becomes a time-like variable, the Jacobian-like additional factors are required since $\vorb$ does not evolve linearly. 
We have made the approximation that the hardening rate $\frac{\dd E_0}{\dd t} \approx E_0\Delta_a/\tenc$.
Use of the Newtonian $\Delta_a$ found in Section~\ref{sec: hardening} is valid when PN effects during interactions are subdominant, which is generally the case when $\tenc<\tgw$.

For the purposes of this calculation, we define three cause-specific hazards:
\begin{gather}
    h_{\rm ej} = \rate{\rm ej} (\vorb' | \theta) \left|\frac{\dd E_0}{\dd t}\right|^{-1} \frac{\dd E_0}{\dd \vorb'} \\
    h_{\rm cap} = \rate{\rm cap} (\vorb' | \theta) \left|\frac{\dd E_0}{\dd t}\right|^{-1} \frac{\dd E_0}{\dd \vorb'} \\
    h_{\rm insp} = \rate{\rm insp} (\vorb' | \theta) \left|\frac{\dd E_0}{\dd t}\right|^{-1} \frac{\dd E_0}{\dd \vorb'}.
    \label{eq:cause_hazards}
\end{gather}
Here we distinguish between binaries which merge between interactions, referred to as {\it inspirals}, and binaries which merge during interactions, referred to as {\it captures} as before. 
For the first two, we compute the rate using the Newtonian $\prob{ej}$ (Eq.~\ref{eq:hillfit}) and post-Newtonian $\prob{cap}$ (Eq.~\ref{eq:p_cap}) fits described above.
We are justified in using the Newtonian form of $\prob{ej}$ because we find the post-Newtonian corrections are only significant when $\tenc>\tgw$ for all $\theta$ considered.
For the third, we must construct the rate by hand since it does not follow from scattering experiments.
After a given binary--single interaction, all binaries will have a new eccentricity drawn from a thermal distribution. 
Some fraction $f_{\rm insp}$ will have a lifetime shorter than the mean time between encounters, which we compute by finding $e_0$ such that $\tenc/\tgw = 1$ and integrating over the thermal distribution.
We take $f_{\rm insp}$ as $\prob{insp}$ to compute $\rate{insp}$.

Then, we sum the cause-specific hazards to find the total hazard $h(\vorb|\theta)$ (Eq.~\ref{eq:totalhazard}) and integrate to obtain the survival function (Eq.~\ref{eq:hazard}),
\begin{equation}
    S(\vorb | \theta) = \exp\left[-\int_{v_{\rm min}}^{\vorb} \dd \vorb' h(\vorb|\theta) \right].
\end{equation}
From this the cause-specific incidence functions $F_C$ follow from Eq.~(\ref{eq:cumulativeincidence}):
\begin{equation}
    F_C(\vorb | \theta) = \exp\left[-\int_{v_{\rm min}}^{\vorb} \dd \vorb' h_C(\vorb|\theta)S(\vorb|\theta) \right] .
\end{equation}

We show an example outcome of this calculation in Figure~\ref{fig:cumulative_ex}. 
In this example, we consider a $300\,\msun$ IMBH encountering a population of $90\,\msun$ BHs with $\vinf = 20$ km/s and $\nfid$.
We compute $F_C$ for the three different outcomes and stack the results to find the total cumulative distribution function $F$.
For convenience, we fill in the the space below each line to correspond to the total probability of each outcome, with blue, orange, and green corresponding to ejection, capture, and inspiral, respectively.
The x-axis is a time-like variable, in this case the hardness, so that the value of each $F_C$ corresponds to the given outcomes occurring before that point in time. 
Here, the binaries have all been terminated by one of the three mechanisms by roughly $\vorb/\vinf=60$, at which point all three curves plateau at their maximum value, which represents the integrated probability of the given outcome.
Here, we can see that the majority of binaries~($59\%$) terminate through inspiral, followed by those which are dynamically ejected~($23\%$) and those merge via capture during interactions~($18\%$).
Binaries which are ejected are mostly ejected between $\vorb/\vinf=20$ and $\vorb/\vinf=40$.

In Figure~\ref{fig:f_ej}, we show $F_{\rm ej}$ as a function of $\vinf$ for different assumed parameters. 
The blue, orange, green, and red lines correspond to $q=0.6, 0.3, 0.2, 0.1$ models, and the different line styles correspond to $m_{00} = 100\,\msun$ (solid), $m_{00} = 300\,\msun$ (dashed), and $m_{00} = 1000\,\msun$ (dotted).
The columns each assume a different potential and the rows a different central $n$.
The range of $\vinf$ chosen roughly corresponds to observed $\vesc$ for GCs and nuclear star clusters \citep{AntoniniF+RasioF2016}.
We find that that as $\vinf$ grows (equivalently, as the cluster mass grows), the fraction of binaries which are dynamically ejected goes down, in agreement with expectations from scaling arguments (Eq.~\ref{eq:tenc_limit}) and Monte Carlo simulations \citep[e.g.,][]{AM+2025}.
$F_{\rm ej}$ is roughly unity at low $\vinf$ but quickly transitions to a power-law, with the start of the transition depending on the assumed parameters.
The largest effect comes from changing $q$, so that it is much harder to eject binaries at low $q$.
For example, in the top left panel (Plummer, $n=10^5\,{\rm pc}^{-3}$), at $\vinf=10$ km/s, the probability of ejecting a $100\,\msun$ BH is roughly $20\%$ for $q=0.2$, but falls to $1\%$ for $q=0.1$.
Changes in the other parameters result in more modest changes.
For example, increasing the mass by a factor of $10$ corresponds to a similar increase in ejection probability in the power law region of the function.
Increasing $n$ by a factor of $10$ results in about a factor of $3$ increase in ejection probability, as this increases the rate of interactions and decreases the window for an isolated inspiral.
This weak dependence on $m_{00}$ and $n$ arise from the similarly weak dependence of $\tgw$ on these parameters.
Changing the assumed potential has the smallest effect on the probability.
Thus, we can conclude that, for the mass ratios shown, dynamical ejection is the most likely outcome at very low $\vinf\sim1$ km/s.
At intermediate $\vinf$, IMBHs may avoid dynamical ejection as long as they are sufficiently larger than the other BHs in the cluster.
At high $\vinf$, dynamical ejection becomes essentially impossible, as expected.

\section{Discussion}
\label{sec:discussion}

\subsection{Gravitational-wave Recoil}

\begin{figure*}
    \centering
    \includegraphics[width=0.92\textwidth]{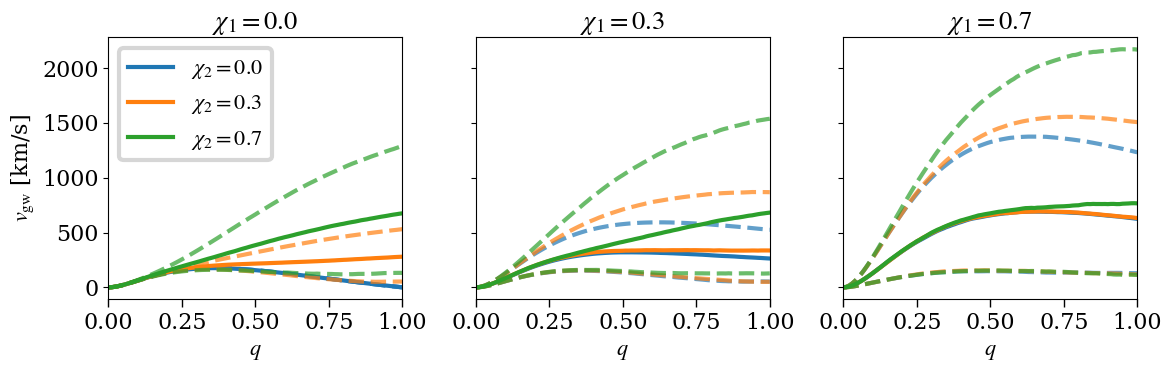}
    \caption{
    Magnitude of $\vgw$ as a function of $q$ for different combinations of spin for the primary and secondary.
    In order to produce the curves shown, we randomize the orientation of both spin vectors isotropically with respect to an arbitrary reference plane $4000$ times for each set of curves and compute the resulting $\vgw$ magnitude using fits to numerical relativity simulations \citep{Campanelli+2007, Gonzalez+2007, Lousto+Zlochower2008, Lousto+2012, Lousto+Zlochower2013}.
    The solid lines show the median recoil magnitude and the dashed lines show the $90\%$ region.
    From left to right, we set $\chi_1$ to $0$, $0.3$, and $0.7$ in the different panels.
    Within each panel, we set the spin of $\chi_2$ to $0$ (blue), $0.3$ (orange), and $0.7$ (green), respectively.  
    }
    \label{fig:vgw}
\end{figure*}

The final ingredient in our model is gravitational-wave recoil from the merger of two BHs.
It is simple to incorporate in our model since the magnitude of the recoil kick $\vgw$ depends on only three dimensionless quantities, $q, \chi_1, \chi_2$, as well as the relative orientations of the BH spins.
The magnitude of $\vgw$ as a function of $\chi_1$ and $\chi_2$, computed from fits to numerical relativity simulations \citep{Campanelli+2007, Lousto+Zlochower2008, Lousto+2010, Lousto+2012}, is shown in Figure~\ref{fig:vgw}.

Previous studies found that these kicks eject the merger products except in the most massive clusters when the natal spin is sufficiently high ($\chi\gtrsim0.5$) \citep[e.g.,][]{AntoniniF+RasioF2016, FragioneG+LoebA2021}.
However, if BHs born from stellar collapse have a negligible natal spin \citep{FullerJ+MaL2019}, then $\vgw$ only depends on $q$, which corresponds to the blue curve in the leftmost panel.
In this case, the maximum value of $\vgw\approx175$ km/s occurs for $q\approx0.35$, which is high enough to eject merger products from the most massive MW GCs.
However, the assumption of negligible natal spin should be taken as the most optimistic estimate.
Remnants of BBH mergers will be born with some spin due to angular momentum conservation.
For example, for a merger between two roughly equal-mass BHs, the spin of the remnant is about $\chi=0.7$.
Moreover, it is possible for BHs to be spun up through accretion following collisions with stars \citep{KirogluF+2025a,KirogluF+2025b,KirogluF+2025c}.
In principle, this means that IMBHs which escape dynamical ejection are still at risk from being ejected by recoil kicks even in the most massive clusters, though this is also a strong function of $q$.

\subsection{Other Escape Mechanisms}

We have reviewed the two dominant mechanisms for ejection of IMBHs.
However, there are other ways by which objects may escape clusters.
All the potential escape mechanisms were studied in detail by \citet{WeatherfordNC+2023}, which we summarize below to discuss the implications for our model.

Escape from clusters can be divided into two categories.
The first, evaporation, is characterized by processes governed by the bulk evolution of the cluster on timescales much longer than the dynamical time.
For example, individual objects may diffuse via $2$-body relaxation to speeds above $\vesc$, which steadily decreases as the cluster loses mass through both dynamical mechanisms and stellar evolution. Furthermore, the strength of the tidal potential changes as the cluster orbits within the host galactic potential, thus potentially lowering the threshold for escaping. These evaporative processes are not expected to play a significant role in the fate of IMBHs, as these processes are most relevant for objects outside the cluster core.
Dynamical friction, on the other hand, causes the orbits of heavy objects of mass $M$ to decay into the center of the cluster on a timescale $t_{\rm df} = \langle m \rangle t_{\rm relx}$, where $\langle m\rangle$ is the average stellar mass and $t_{\rm relx}$ the relaxation time.
For a typical cluster, $t_{\rm df} \sim 1$ Myr.
Additionally, an object that is only marginally unbound can take many relaxation timescales to actually escape, during which time the object can be scattered back into a lower-energy orbit.
While it is possible that IMBH binaries which we considered marginally bound in our scattering experiments could escape through evaporative mechanisms, this correction is small and only relevant for low $\vorb/\vinf$ when escape is already so unlikely as to be negligible. 
Thus, it is unlikely that this would have a noticeable effect on $F_{\rm ej}$ calculated in Fig.~\ref{fig:f_ej}.

On the other hand, ejection is expected to dominate in the core of the cluster, which was the environment we considered in this work.
Escape by ejection is characterized by single, impulsive events that impart a recoil on objects greater than $\vesc$.
This includes natal supernova kicks, strong gravitational interactions, and collision-induced recoil kicks.
Strong gravitational interactions include, in addition to the binary--single interactions studied here, strong 2-~and~3-body encounters between unbound singles and higher order few-body encounters.
By construction, we have neglected supernova kicks in our model by assuming the presence of a population of BHs. 
BH natal kicks are highly uncertain; while observations of BH X-ray binaries suggest some BHs receive natal kicks as high as $100\,\mathrm{km/s}$, many may receive kicks that are much lower \citep{RepettoS+2017}.
Moreover, the observational and theoretical support for large populations of BHs in MW GCs support low natal kicks \citep[e.g., ][]{WeatherfordN+2020}.
We similarly neglect the effects of 2-body and 3-body encounters between singles as IMBHs are expected to efficiently exchange into binaries \citep[e.g., ][]{SS+PES1993, HeggieDC+1996}.
Moreover, while \citet{WeatherfordNC+2023} found that 3-body binary formation dominated the rate of high speed escapers, these were typically small single stars encountering much larger objects.
Indeed, \citet{AtallahD+2024} found that binaries containing BHs formed via 3-body encounters rarely have a recoil above the $\vesc$ assumed in our models.
Similar reason holds for strong 2-body encounters; typically, an IMBH will impart a large kick onto a much smaller object, but the reverse is energetically not possible.

However, we must justify our suppression of binary--binary and higher order few-body encounters.
The density of binaries is suppressed as soft BBHs tend to be ionized by the few (or single) harder BBH through direct ionization, triple formation, or gravitational-wave capture during a resonant interaction \citep{DMP+MG2024,DMP+2025}. 
Higher-order multiple systems are similarly expected to be quickly ionized in dense environments.
Moreover, while the additional binding energy of the second binary can, in principle, result in higher $\vkick$ compared to a single object of the same mass, the energy budget of the interaction will still be dominated by the harder and more massive IMBH binary.
Thus, we would not expect that the presence of a small number of BBHs to significantly affect either the rate of hardening or the rate of ejections. 
Indeed, \citet{WeatherfordNC+2023} reports that while the binary--binary mechanism results in a marginally higher median recoil speed, the rate of ejections is dominated by the binary--single mechanism.

In principle, it is possible that this assumption causes us to underpredict the number of mergers during interactions, as it has been shown that binary--binary interactions are responsible for a very large fraction of captures in the equal--mass case \citep{ZevinM+2019}.
Similarly, secular evolution in triples formed by binary--binary interactions may also contribute to the merger rate, though the rate is very small compared to the total \citep{MartinezM+2020}.
This may be important when considering the role of gravitational-wave recoil in ejecting IMBHs, even if this channel constitutes only a fraction of the total merger rate.
Nevertheless, work in progress is underway studying the role of binary-binary encounters at early times involving two or more IMBHs (Zhou et al., in prep.).

\subsection{Applications for Realistic Stellar Systems}

The model presented in the previous section is highly idealized in comparison to realistic dense stellar environments.
First, in realistic stellar systems, the mass distribution for BHs arises from a stellar IMF which is subsequently modified by collisions during dynamical interactions before the massive stars collapse into BHs.
Second, $\sigdisp$ is not static, as is assumed in our model, but decreases from its initial value over time as the cluster loses mass through dynamical processes, so that it becomes easier to eject IMBHs over time.
Nevertheless, it is expected that the largest BHs that form in a dense stellar environment will begin encountering each other due to mass segregation on timescales much faster than a relaxation time \citep[within the first Gyr; see, e.g., ][]{EGP+2022}.
Therefore, the fate of IMBHs formed in these environments depends on the initial $\sigdisp$ and $\vesc$ and on the initial masses of the most massive BHs.

Previous work using $N$-body modeling studying the formation of IMBHs in dense stellar environments has shown that their formation is strongly dependent on highly uncertain initial conditions.
Systematic exploration of initial conditions with both Monte Carlo simulations \citep{EGP+2021, NW+2021, EGP+2022, EGP+2024, SharmaK+RodriguezCL2025, Khurana2025} and direct $N$-body methods \citep{FROSTCLUST1,FujiiM+2024,VMC+2025, FROSTCLUST2, FROSTCLUST3} have shown that the formation of VMS progenitors of IMBHs, is strongly dependent on both the initial density of massive stars and the assumed stellar metallicity.
A larger density of massive stars promotes a higher rate of stellar collisions prior to BH formation, while low metallicities are required such that stellar winds do not significantly erode the resulting hydrogen envelope.
In these studies, it is shown that the largest IMBHs formed from direct collapse can be in the range of $10^2$ to a few $10^4\,\msun$ depending on initial conditions.
When runaway collisions in extremely dense environments result in the production of a single large VMS, the resulting IMBH with $M\gtrsim10^3\,\msun$, formed from direct collapse reduces the number of BHs through the merger of massive star progenitors \citep{SharmaK+RodriguezCL2025}.
On the other hand, in simulations where the largest IMBH is only a few $10^2\,\msun$, many IMBHs of similar mass and mass gap BHs may form, making retention unlikely in all but the most massive clusters \citep{EGP+2022}.

While the initial properties of dense stellar clusters---such as the IMF, binary fractions, and cluster sizes--- remain highly uncertain, JWST is starting to reveal the properties of young massive clusters (YMCs) at high redshifts.
At redshift $z \sim 6$, NIRCAM imaging reveals YMCs with masses roughly $10^{6-7}\,\msun$ \citep{Vanzella2022b}.
In the Cosmic Gems Arc, five new proto-GC candidates have been found that exhibit stellar surface densities three orders of magnitude higher than young star clusters in the local universe \citep{Adamo2024}. 
Similar observations of stellar populations were recently reported in the Cosmic Spear \citep{COSMICSPEAR}.
More observational and theoretical work is necessary to constrain the initial properties of dense stellar environments in order to connect the clusters observed at these early times to those observed in the present and thus constrain the likelihood of IMBH formation.

\subsection{Implications for IMBH candidates in the MW GCs}

While direct detections of IMBHs in GCs remain elusive, several clues of their presence have begun to emerge. 
\cite{Haberle2024} report the detection of seven fast-moving stars, each exceeding the local inferred escape velocity, at the center of Omega Centauri. 
Alternative explanations for these high velocities, including foreground objects or tight binaries around a stellar mass BH, have been ruled out. 
Their findings therefore strongly suggest the presence of an IMBH with a mass in the range $8,200-50,000\,\msun$ in Omega Centauri, which is theorized to be the stripped nucleus of a dwarf galaxy. 
Recently, \texttt{CMC} simulations--incorporating the relevant loss-cone physics for a fixed central IMBH--have modeled the growth of the central IMBH and reproduced observed velocity and brightness profiles \citep{EGP+2025}. 
\citet{EGP+2025} found that models were most consistent with the observed properties of Omega Centauri when the simulation initial conditions included a truncation of the high end of the stellar IMF. 
This happens to be consistent with seed formation from runaway stellar collisions such that only one IMBH (the fixed potential) is initially formed such that it is immune to ejection by the stellar BHs that later form from stellar collapse.

Another cluster of interest is 47 Tucanae, where \cite{Paduano2024} identified a compact central radio source. 
Though the presence of a weakly accreting IMBH with a mass in the range $54 - 6000\, \msun$ can be invoked, an undiscovered millisecond pulsar can't be ruled out. 
Furthermore, dynamical modeling by \citet{DellaCroceA+2024} demonstrates that the observed stellar kinematics of the cluster's central region are incompatible with central IMBH masses greater than $578\,\msun$. 
This more stringent upper limit on the mass is incompatible with our results. 
\citet{YeCS+2022} showed that the present-day observational properties of 47 Tucanae are well matched by a model containing a $\sim2000\,\msun$ BH subcluster and present day $\sigdisp\simeq15\,\mathrm{km/s}$ without requiring the presence of an IMBH.
The best-fit model presented there has an initial number of BHs roughly a factor of a few times higher than the present, corresponding to a similar factor increase in the total BH mass, and $\sigdisp\simeq30-40\,\mathrm{km/s}$.
While ejection at early times can be avoided by assuming that no IMBHs of comparable mass (say, $q\gtrsim0.2-0.3$) were simultaneously formed, a $600\,\msun$ IMBH would need to be roughly $10$ times larger than the next most massive BHs to avoid ejection at times closer to the present.
This is in tension with expectations from simulations.
Furthermore, even a modest amount of spin would make such an IMBH liable to ejection from the cluster, even with $q\simeq0.1$.

\subsection{Gravitational Wave Observations}

Though it may be unlikely to detect low-mass IMBHs through direct observations in the MW GCs, the low-mass IMBHs that are ejected in binaries through dynamics are still promising as potential multiband gravitational-wave sources with current and future gravitational wave detectors such as LISA, Cosmic Explorer, and the Einstein Telescope \citep[see][for a detailed discussion]{JaniK+2020}. 
Detailed rate estimates for different instruments are outside the scope of this work, but we can make qualitative comments in the context of our results. 

Analysis of the population properties of GWTC-4 data has shown that the differential merger rate for the underlying astrophysical population is consistent with a power law with index $\beta_q=1.2^{+1.2}_{-1.0}$, so that mergers with $q\simeq1$ are $1000$ times more common than mergers with $q\simeq0.1$ \citep{GWTC4population}.
Our results suggest that, since IMBHs with comparable masses are more likely to be ejected, they will tend to merge at redshifts near $z=0$, which is consistent with the reported astrophysical population properties.
On the other hand, IMBH binaries with large mass asymmetry tend to merge within their host cluster, such that these mergers are more likely to happen at early times.
Thus, it is possible that IMBH mergers with $q\lesssim0.2$ may be more common at higher redshift.
This is especially true if some of these IMBHs are retained following the merger, as they will merge efficiently with stellar-mass BHs \citep{EGP+2025}.

\section{Summary}
\label{sec:conclusion}

We have studied the relative importance of dynamical recoil and gravitational-wave recoil for ejecting IMBHs from star clusters.
For this end, we have presented the results of both Newtonian and post-Newtonian scattering experiments of intermediate mass ratios.
From our scattering experiments, we provided fits to the branching ratios for ejection $\prob{ej}$ and gravitational-wave capture $\prob{cap}$, as well as the per-encounter hardening rate $\langle\Delta\rangle$ for binary--single interactions where $m_{00}>m_{01}=m_1$ for $q=0.01-0.6$.
Using these fits, we created a toy model for IMBH binary evolution in dense stellar environments and use methods borrowed from survival analysis to compute the probability of ejection for an IMBH-BH binary through dynamical interactions.
We find that the probability of ejection for a given IMBH binary is most sensitive to the cluster velocity dispersion ($\sigdisp$) and the mass ratio of the IMBH with respect to other BHs in the cluster, with dynamical ejection being dominant at low $\sigdisp$ and improbable for high $\sigdisp$.
However, IMBHs which are immune to dynamical ejection may still be at risk for ejection by gravitational wave recoil depending on assumption about the natal spin.
We discuss the implications of our toy model for observations of IMBHs and find that, while direct detection of IMBHs with masses of order $10^2\,\msun$ is unlikely in present-day clusters, it is possible that future gravitational wave detectors might detect the merger of IMBH binaries in this mass range with asymmetric mass ratios.

\section{Acknowledgments}

We thank Christopher O'Connor, Sanaea Rose, Dany Atallah, Giacomo Fragione, Yonadav Barry Ginat, Nicholas Stone, Aaron Tierney, and Emiko Kranz for helpful discussions and comments.
This work was supported by NSF Grants AST-2108624 and AST-2511543 at CIERA, and used computing resources funded by NSF Grants PHY-1726951 and PHY-2406802. 
Our research was supported in part through the computational resources and staff contributions provided for the Quest high performance computing facility at Northwestern University, which is jointly supported by the Office of the Provost, the Office for Research, and Northwestern University Information Technology.
Support for E.G.P.\ was provided by the NSF Graduate Research Fellowship Program under Grant DGE-2234667.

\software{
\texttt{Fewbody} \citep{fewbody}, \texttt{NumPy} \citep{numpy},
\texttt{SciPy} \citep{scipy},
\texttt{pandas} \citep{pandas2010, pandas2020},
\texttt{Matplotlib} \citep{matplotlib}
}

\bibliography{refs}{}
\bibliographystyle{aasjournal}

\appendix

\section{Central Escape Speed in King Profile}
\label{sec:king}
\begin{figure}
    \centering
    \includegraphics[width=0.46\textwidth]{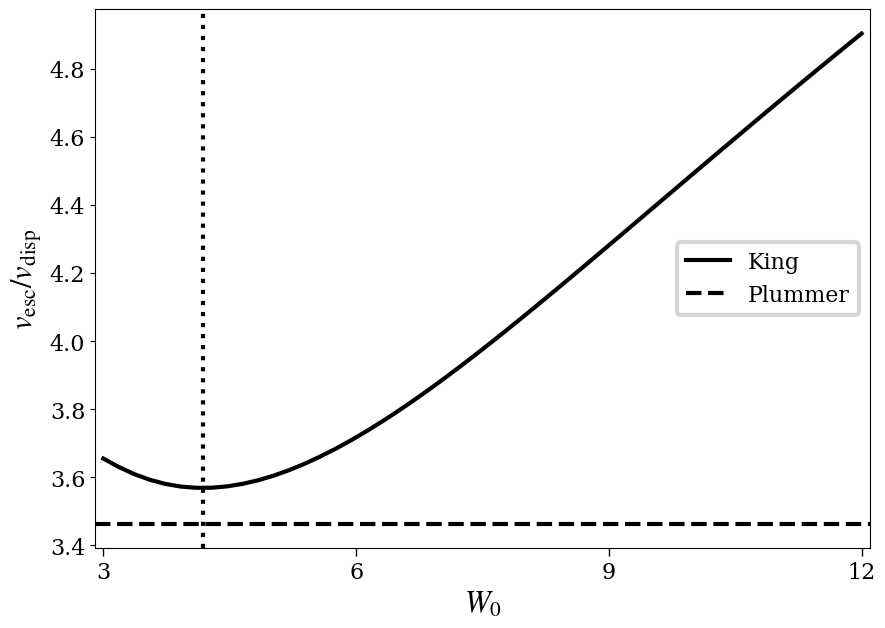}
    \caption{
    Ratio of escape speed from the center of the cluster to velocity dispersion for a King profile with different values of $W_0$. 
    The dashed black line showed the corresponding quantity for a Plummer profile.
    }
    \label{fig:kingesc}
\end{figure}

The King profile is characterized by the distribution function
\begin{equation*}
    f(\mathcal{E}) = \rho_0 (2\pi\sigma^2)^{-3/2} (e^{\mathcal{E}/\sigma^2} - 1)
\end{equation*}
when the relative energy $\mathcal{E} = \Psi - v^2/2 > 0$ and is equal to $0$ otherwise, $\Psi(r) = \Phi(r_t) - \Phi(r)$ is the relative potential, $r_t$ is the tidal radius, and $\rho_0$ is the central density. 
Note that the parameter $\sigma$, which characterizes the thermal energy of the objects in the cluster, is closely related but not equivalent to the physical velocity dispersion $\sigdisp$. 
Integrating the distribution function over all velocities, we find \citep[e.g., ][]{BTtextbook}
\begin{equation*}
    \rho(\Psi) = \rho_0\left[e^{\Psi/\sigma^2} {\rm erf}\left({\sqrt{\frac{\Psi}{\sigma^2}}}\right) - \sqrt{\frac{4\Psi}{\pi\sigma^2}} \left( 1 + \frac{2\Psi}{3\sigma^2}\right) \right]
\end{equation*}
Now, from here, one may wish to find the functional form of the relative potential $\Psi(r)$, which would require integrating the equation of Poisson, which would allow one to find a functional form for $\rho(r)$. 
However, it is clear from the form of $\rho(\Psi)$ that the family of King models can be described by a single parameter $W_0 = \Psi(0)/\sigma^2$. 
This value sets the shape of the density--potential pair. 
Choosing a radial scale, such as the King (core) radius
\begin{equation*}
    r_c = \sqrt{\frac{9\sigma^2}{4\pi G\rho_0}}
\end{equation*}
sets the scale of the system. 
Finally, setting either a central density $\rho_0$ or a total cluster mass $M_{\rm tot}$ allows one to find a dimensionful value of the parameter $\sigma$, which completely determines the parameters of the cluster model. 
The escape velocity from the cluster core is
\begin{equation}
    \vesc(0) = \sqrt{2\Psi(0)} = \sqrt{2 W_0}\sigma.
\end{equation}
The velocity dispersion $\sigdisp$ can be found by integrating the appropriate moment of the distribution function \citep{HD1987}: 
\begin{equation*}
    \sigma^2(\Psi) = \frac{1}{\rho(\Psi)} \int_0^\Psi \rho(\Psi') \dd{\Psi'}.
\end{equation*}
Thus we find that $\sigdisp$ is related to $\sigma$ by the relation
\begin{equation}
\begin{split}
    \sigdisp^2(\Psi) = \sigma^2 \Biggl( e^{\Psi/\sigma^2} {\rm erf} \left({\sqrt{\frac{\Psi}{\sigma^2}}}\right) - \frac{2}{\sqrt{\pi}} \left( \frac{\Psi}{\sigma^2} \right)^{1/2} \\- \frac{4}{3\sqrt{\pi}} \left( \frac{\Psi}{\sigma^2} \right)^{3/2} - \frac{8}{15\sqrt{\pi}} \left( \frac{\Psi}{\sigma^2} \right)^{5/2} \Biggr) \\
    \left[e^{\Psi/\sigma^2} {\rm erf} \left({\sqrt{\frac{\Psi}{\sigma^2}}}\right) - \sqrt{\frac{4\Psi}{\pi\sigma^2}} \left( 1 + \frac{2\Psi}{3\sigma^2}\right)\right]^{-1}
\end{split}
\end{equation}

In Figure~\ref{fig:kingesc}, we show the ratio of the central values $\vesc/\sigdisp$ as a function of $W_0$ (solid) for the King model compared to that for a Plummer model (dashed). 
The value of the ratio $\vesc/\sigdisp$ for King models is higher than that of the Plummer profile within the range of interest by an order unity factor and is closest for $W_0 \approx 4.2$.

\section{Survival Analysis and Competing Risk Analysis}
\label{sec:survivalanalysis}
\setcounter{equation}{0}
In this study, we make use of the results of survival analysis, which are used widely in other fields concerned with system failure, e.g. human mortality rates, failure of mechanical components, etc. 
In these fields, the below methods are standard and many references can be easily found, but we refer the reader to \citet{SurvivalAnalysis} for an extremely mathematically rigorous treatment of this subject.

The rate of ejections $\rate{ej}$ shares properties with what is called the hazard function $h(t)$, defined as
\begin{equation}
    h(t) = \lim_{\Delta t \to 0} \frac{P\!\left(t \le T < t+\Delta t \mid T \ge t\right)}{\Delta t}\,,
    \label{eq:hazarddef}
\end{equation}
where $T$ is the true failure time.
This can be understood as the instantaneous rate of failure of a system at time $t$ given that the system has survived until $t$. 
By manipulating Eq.~(\ref{eq:hazarddef}), it can be shown that $h(t)$ is related to the probability distribution function of failure times $f(t)$ and the survival function $S(t) = 1 - F(t)$ by 
\begin{equation}
    h(t) = \frac{f(t)}{S(t)} = -\frac{\dd}{\dd t}\ln S(t).
    \label{eq:hazard}
\end{equation}
The significance of these functions for our results is as follows: if we can construct $h(\vorb)$ from $\rate{ej}$, then $S(\vorb)$ corresponds to the fraction of systems that have survived ejection at a given $\vorb$. 
Then, $f(\vorb)$ corresponds to the distribution with which binaries are ejected, which can be transformed into, e.g., a delay time distribution of gravitational-wave merger events.

However, just as there are multiple competing causes of mortality, the binaries too have different end states. 
Thus, we need to use competing risk analysis, which generalizes the above for cause-specific hazard functions $h_c$.
The total hazard is
\begin{equation}
    h(t)=\sum_c h_c(t)
    \label{eq:totalhazard}
\end{equation}
such that, if the intrinsic hazards are independent of each other, the incidence probability for a single failure mode is
\begin{equation}
    I_c (t) = S(t)h_c(t).
\end{equation}
Note that the incidence probability is \textit{not} a probability distribution, which can be seen from Eq.~(\ref{eq:hazard}).
The corresponding analog of $F(t)$ is the cumulative incidence function (commonly abbreviated as CIF), defined as
\begin{equation}
    F_c(t) = \int_0^t h_c(t')S(t')\dd t'.
    \label{eq:cumulativeincidence}
\end{equation}
$F_c(t)$ is the cause-specific failure before time $t$, taking into account the condition that competing failure causes have not taken place. 
This function has the property $F(t) = \sum_c F_c(t)$ due to the linearity of the previous definitions.

\end{document}